\documentclass[a4paper,prb,showpacs,preprintnumbers,amsmath,amssymb]{revtex4}
\usepackage{hyperref}
\allowdisplaybreaks[4]

\newcommand{\alphabar}{\mkern 3.5mu\overline{\mkern -3.5mu\alpha\mkern -2.5mu}\mkern 2.5mu}
\newcommand{\betabar}{\mkern 3mu\overline{\mkern -3mu\beta\mkern -1mu}\mkern 1mu}
\newcommand{\sigmabar}{\mkern 3mu\overline{\mkern -3mu\sigma\mkern -2mu}\mkern 2mu}

\begin {document}

\title{Effective Hamiltonian for a Half-filled Asymmetric
Ionic Hubbard Chain with Alternating On-site Interaction}
\author{Inna Grusha$^{a}$, Mikheil Menteshashvili$^{b}$
and George I. Japaridze$^{a,c}$}

\affiliation{\vspace{2mm}\\
$^{a}$\,Faculty of Natural Sciences and Engineering, Ilia State
University, Cholokashvili Avenue 3-5, 0162 Tbilisi, Georgia\\
$^{b}$\,Department of Physics, University of Fribourg, Chemin du
Mus\'ee 3, CH-1700 Fribourg, Switzerland\\
$^{c}$\,Andronikashvili Institute of Physics, Tamarashvili str.~6,
 0177 Tbilisi, Georgia}

\begin{abstract}

We derive an effective spin Hamiltonian for the one-dimensional
half-filled asymmetric ionic Hubbard model with alternating on-site
interaction in the limit of strong repulsion. It is shown that the
effective Hamiltonian is that of a spin $S=1/2$ anisotropic $XXZ$
Heisenberg chain with alternating next-nearest-neighbor and
three-spin couplings in the presence of a uniform and a staggered
magnetic field.
\end{abstract}

\pacs{71.27.+a Strongly correlated electron systems; heavy fermions,
75.10.Jm Quantized spin models}

\maketitle

\section{Introduction} \label{intro}

During the last decades the correlation-induced metal-insulator
(Mott) transition has been one of the challenging problems in
condensed matter physics. \cite{ReviewImada} In most cases the
translational symmetry is broken in the Mott insulator. \cite{Lee} A
notable exception is the one-dimensional repulsive ($U>0$) Hubbard
model \cite{HubbardBook}
%%%%%%%%%%%%%%%%%%%%%%%%%%%%%%%%%%%%%%%%%%%%%%%%%%%%%%%%%%%%%%%%%
\begin{equation}\label{HamiltonianHubMod} {\cal H}_{Hub} =
t\sum_{n}\sum_{\alpha=\uparrow,\downarrow}\big(c^{\dag}_{n\alpha}
c^{\phantom{\dag}}_{n+1,\alpha} +
c^{\dag}_{n+1,\alpha}c^{\phantom{\dag}}_{n\alpha} \big)
 + U \sum_{n}{\hat \rho}^{\phantom{\dag}}_{n\uparrow} {\hat
\rho}^{\phantom{\dag}}_{n\downarrow}
\end{equation}
%%%%%%%%%%%%%%%%%%%%%%%%%%%%%%%%%%%%%%%%%%%%%%%%%%%%%%%%%%%%%%%%%
at half-filling, where the dynamical generation of a charge gap
is not accompanied by the breaking of a discrete symmetry. \cite{LiebWu}
In equation (\ref{HamiltonianHubMod}) we have used standard notation,
namely $c^{\dag}_{n\alpha}$ ($c^{\phantom{\dag}}_{n\alpha}$) for
electron creation (annihilation) operators and
${\hat\rho^{\phantom{\dag}}}_{n\alpha}=c^{\dag}_{n\alpha}
c^{\phantom{\dag}}_{n\alpha}$ for the particle density at site
$n$ with spin projection $\alpha$. The exact solution of the model
(\ref{HamiltonianHubMod}) in the case of a half-filled band reveals
that the ground state is uniform, with exponentially decaying
density correlations. \cite{EsslerFrahm99} At the same time, spin
excitations are gapless and thus magnetic correlations decay only
algebraically. \cite{FrahmKorepin90} This is readily understood in the
large-$U$ limit: indeed, for $U \gg |t|$ the infrared behavior of
the model (\ref{HamiltonianHubMod}) at half-filling
is fully described by the $SU(2)$-symmetric spin $S=1/2$
Heisenberg Hamiltonian
%%%%%%%%%%%%%%%%%%%%%%%%%%%%%%%%%%%%%%%%%%%%%%%%%%%%%%%%%%%%%%%%%
\begin{equation}
{\cal H}_{\it Heis}= J\sum_{n} {\bf S}_{n} \cdot {\bf S}_{n+1} +
J^{\prime}\sum_{n} {\bf S}_{n} \cdot {\bf S}_{n+2}\, ,
\label{J-J'-HeisChain}
\end{equation}
%%%%%%%%%%%%%%%%%%%%%%%%%%%%%%%%%%%%%%%%%%%%%%%%%%%%%%%%%%%%%%%%%
where $J=4t^{2}/U -16t^{4}/U^{3}$ and $J^{\prime}=4t^{4}/U^{3}$ up
to the fourth-order terms in $t/U$. \cite{Bulaevskii67,Takahashi77} Since the
condition $|t| \ll U$ implies that the frustration is weak
$J^{\prime} \ll J$, the next-nearest exchange is irrelevant and the
low-energy behavior of the initial electron system is governed
by the standard isotropic Heisenberg Hamiltonian ${\cal H}_{\it
Heis}= J\sum_{n} {\bf S}_{n} \cdot {\bf S}_{n+1}$. Elegant techniques
have been developed for calculating higher-order corrections to
the Hamiltonian (\ref{J-J'-HeisChain}). These terms are also
irrelevant and leave the featureless character of the ground state
intact. \cite{MacDonald88,Stein97,Datta99}

The spin sector may even remain translationally invariant in the case
of an explicitly broken translational invariance of the electronic
Hamiltonian. For example, let us consider a scenario where two types
of atoms are located respectively on even and odd sites of the lattice,
with different on-site energies and/or different on-site couplings for
the electrons. The Hamiltonian of such an extended version of the
Hubbard model is given by
%%%%%%%%%%%%%%%%%%%%%%%%%%%%%%%%%%%%%%%%%%%%%%%%%
\begin{equation}\label{ExtHubbardHamiltonian-1}
{\cal H} =
t\sum_{n,\alpha}\big(c^{\dag}_{n\alpha}
c^{\phantom{\dag}}_{n+1,\alpha} +
c^{\dag}_{n+1,\alpha}c^{\phantom{\dag}}_{n\alpha} \big)
+\,U \sum_{n}\big(\,1+(-1)^{n}\delta\,\big){\hat
\rho}^{\phantom{\dag}}_{n\uparrow} {\hat
\rho}^{\phantom{\dag}}_{n\downarrow}
+\frac{\Delta}{2}\sum_{n}\sum_{\alpha} (-1)^{n}\,{\hat
\rho}^{\phantom{\dag}}_{n\alpha} \, ,
\end{equation}
%%%%%%%%%%%%%%%%%%%%%%%%%%%%%%%%%%%%%%%%%%%%%%%%
where $0\leq \delta,\, \Delta/U \ll 1$. It possesses spin $SU(2)$
symmetry, but the translational symmetry has been reduced due to the
doubling of the unit cell. At $\delta=0$ and $\Delta \neq 0$ this
Hamiltonian corresponds to the ionic Hubbard model (IHM),
\cite{Nagaosa86} where electrons on even and odd sites have
different on-site energies $\pm \Delta/2$, while at $\Delta=0$ and
$\delta \neq 0$ equation (\ref{ExtHubbardHamiltonian-1})
represents the alternating-$U$ Hubbard model, \cite{AHM93} where the
electrons experience different on-site interactions on even and odd
sites.

At $U$ = 0 the half-filled ionic Hubbard model describes a regular
band insulator with equal charge and spin gaps and a long-range
ordered (LRO) charge-density-wave (CDW) in the ground state. With
increasing $U$ the system undergoes two phase transitions, a first
one at $U=U_{c1}$ from the CDW-insulator to a LRO dimerized insulator,
and a second one at $U=U_{c2}>U_{c1}$ from the dimerized phase to a
strongly correlated (Mott) insulator. \cite{Fabrizio99} At $U=U_{c2}$
the spin gap vanishes and the low-energy behavior of the system for $U
> U_{c2}$ is again described by the Heisenberg Hamiltonian
(\ref{J-J'-HeisChain}), with the difference that the spin exchange
parameters $J$ and $J^{\prime}$ now weakly depend on $\Delta$. The
broken translational symmetry of the model manifests itself only in
the charge degrees of freedom via the presence of a LRO CDW pattern
which persists even in the limit of strong repulsion, with the
amplitude approaching zero at $U\rightarrow\infty$. \cite{KSJB03}

The weak-coupling renormalization group analysis of the repulsive
alternating-$U$ Hubbard model ($\Delta = 0$ and $U(1 \pm \delta)>0$)
shows a qualitatively similar low-energy behavior at half-filling as
the usual Hubbard model. Scattering processes arising from the
alternating part of the interaction, which are relevant in the
commensurates case of 1/4- and 3/4-filled bands, \cite{AHM93} are
irrelevant at 1/2-filling where the properties of the system are
governed by the uniform part of the interaction. In the limit of
strong on-site repulsion ($U \gg |t|$), the infrared behavior of the
alternating-$U$ Hubbard model is once again described by the
Heisenberg Hamiltonian (\ref{J-J'-HeisChain}), but with a slight
modification -- an alternating next-nearest-neighbor (nnn) exchange
$\sum_{n}\big[J^{\prime}_{0}+(-1)^n J^{\prime}_{1}\big]\, {\bf
S}_{n} \cdot {\bf S}_{n+2}\,$. \cite{AHM04} Numerical and
analytical studies of the Heisenberg chain with alternating nnn
exchange show that in the pertinent case of weak frustration
($J^{\prime}\ll J$), the alternation of $J^{\prime}$ is irrelevant
and the infrared behavior of the model is fully described by the
standard Heisenberg model with nearest-neighbor exchange.
\cite{SorellaParola02-SenSarkar} Thus, even though the Hamiltonian
(\ref{ExtHubbardHamiltonian-1}) describes a fermion system on a
lattice with broken translational symmetry, the information about
the unit cell doubling at half-filling is fully accommodated within
the high-energy degrees of freedom; the low-energy behavior of the
system is described by a translationally invariant, isotropic spin
Hamiltonian. It has to be noted that the above conclusion does not
remain valid in the presence of bond alternation, i.e. if the
hopping amplitude $t$ is replaced by $t_{0} + (-1)^n t_{1}$. In this
case one obtains that in the strong-coupling limit at half-filling
the effective Hamiltonian, still given by the Heisenberg model,
contains an alternating nearest-neighbor exchange
$\sum_{n}\big[J_{0} +(-1)^n J_{1}\big]\,{\bf S}_{n} \cdot {\bf
S}_{n+1}$, which leads to the spin-Peierls instability with gapped
spin excitations. \cite{Cross-Fisher-79}

We now turn our attention to a model having full translational
symmetry, but explicitly broken spin $SU(2)$ symmetry, the so-called
spin-asymmetric Hubbard Hamiltonian
%%%%%%%%%%%%%%%%%%%%%%%%%%%%%%%%%%%%%%%%%%%%%%%%%%%%%%%%%%%%%%%%%
\begin{equation}\label{SpinAsmHubbardHamilt}
{\cal H} = \sum_{n,\alpha}t_{\alpha}\big(c^{\dag}_{n\alpha}
c^{\phantom{\dag}}_{n+1,\alpha} +
c^{\dag}_{n+1,\alpha}c^{\phantom{\dag}}_{n\alpha} \big)
+ U\sum_{n}{\hat \rho}_{n\uparrow}
{\hat \rho}_{n\downarrow} \, ,
\end{equation}
%%%%%%%%%%%%%%%%%%%%%%%%%%%%%%%%%%%%%%%%%%%%%%%%%%%%%%%%%%%%%%%%%
where the hopping is spin-dependent ($t_{\uparrow}\neq t_{\downarrow}$).
This model, introduced in the early 1990s \cite{Brandt91} to interpolate
between the standard Hubbard model ($t_{\uparrow}=t_{\downarrow}$)
and the Falicov-Kimball model \cite{FKM69} ($t_{\uparrow}>0,t_{\downarrow}
=0$), has been intensively studied during the last two decades.
\cite{Domanski94,Fath95,Domanski96,Stasyuk-Hera-03,Souza-Macedo-06,Wang-Chen-Gu-07}
Away from half-filling the spin-up and spin-down particles are
segregated in the ground state for large enough repulsion, both for
the Falicov-Kimball model \cite{Freericks-Lieb-Ueltschi-02} and for
the Hamiltonian (\ref{SpinAsmHubbardHamilt}) with $t_{\uparrow}
\neq t_{\downarrow} \neq 0$.\cite{Ueltschi04} Therefore the
spin-asymmetric Hubbard model appears to be well suited for studying
transitions between phase-separated and homogeneous states, especially
in one dimension. \cite{Silva-Valencia07,Gu-Fan-Lin-07,Farkasovsky08}

More recently, increased interest in low-dimensional correlated
fermion models with spin-dependent hopping has been triggered by the
fascinating progress in experimental studies of low-dimensional
mixtures of optically trapped ultracold atoms of two different
types, \cite{UCA-Review-1} such as ultracold atoms loaded into
spin-dependent optical lattices \cite{Mandel04,Messer15a} or trapped atoms
of different masses. \cite{Wille-et-al-08,Taglieber-et-al-08} The
great freedom available for generating optical lattices has also
allowed one to play with the lattice geometry and to create bipartite
lattices, which turned out to be a key ingredient for achieving
higher-band condensates, \cite{hemmerich2011,hemmerich2011PRL,olschlager}
coherence control, \cite{Marco-Cristiane-2014} density-wave
dynamics, \cite{trotzky2012} and even graphene-like physics.
\cite{tarruell2012,Messer15b} It has to be emphasized that mixtures of
fermions with different hopping amplitudes naturally appear in
solid-state systems as well, namely when several bands cross the
Fermi surface. This happens for instance in mixed-valence materials,
organic superconductors, \cite{Penc-Solyom-90} small radius nanotubes,
\cite{Carpentier-Orignac-06} and even graphene-based
heterostructures. \cite{Blanter08-Affleck10} However, experiments
with trapped ultracold atoms can actually engineer quantum many-body
states and thus realize models of correlated fermions and bosons
which are not available in usual solid-state structures. \cite{Wilczek04}
Recent theoretical predictions of various unconventional superfluid or
superconducting, \cite{Giamarchi,vanDongen08,Sarma09,Jolicoeur,FF,Boschi,DiLiberto14a}
insulating \cite{Winograd,DiLiberto14b} and magnetic \cite{Farkasovsky12}
phases in such novel systems have further stimulated the interest in the
spin-asymmetric Hubbard model.

The broken $SU(2)$ spin symmetry of the model (\ref{SpinAsmHubbardHamilt})
at $t_{\uparrow} \neq t_{\downarrow}$ is manifestly seen
for a half-filled band in the strong-coupling limit ($U \gg
|t_{\uparrow}|, |t_{\downarrow}|)$, where to leading order the
infrared behavior of the system is described by the anisotropic
$XXZ$ Heisenberg Hamiltonian
%%%%%%%%%%%%%%%%%%%%%%%%%%%%%%%%%%%%%%%%%%%%%%%%%%%%%%%%%
\begin{equation}\label{XXZ-chain-1}
{\cal H} =  J\sum_{n}\left( S^{x}_{n} S^{x}_{n+1} +  S^{y}_{n}
S^{y}_{n+1}+\gamma \, S^{z}_{n} S^{z}_{n+1}\right)\, ,
\end{equation}
%%%%%%%%%%%%%%%%%%%%%%%%%%%%%%%%%%%%%%%%%%%%%%%%%%%%%%%%
with $J=4t_{\uparrow} t_{\downarrow}/U$ and
$\gamma=(t^{2}_{\uparrow}+t^{2}_{\downarrow})/2t_{\uparrow}
t_{\downarrow}$. \cite{Fath95} As the anisotropy parameter
$|\gamma|$ is larger than $1$ for arbitrary $t_{\uparrow} \neq
t_{\downarrow}$, the system has a finite spin gap and long-range
antiferromagnetic order in the ground
state. \cite{Cloizeaux-Godin-65} Nevertheless, the translational
invariance of the initial lattice model (\ref{SpinAsmHubbardHamilt})
is retained by the effective Hamiltonian (\ref{XXZ-chain-1}), even
if the ground state has lower symmetry due to the general phenomenon
of spontaneous symmetry breaking.

In a recent paper we have studied the one-dimensional
spin-asymmetric ionic Hubbard model in the limit of strong on-site
repulsion (for a half-filled band). \cite{Grusha-Japaridze-13} We
have shown that for $t_{\uparrow} \neq t_{\downarrow}$ the doubling
of the unit cell by the alternating ionic potential $\Delta \neq 0$
directly manifests itself in the spin degrees of freedom, and the
effective spin Hamiltonian in the strong-coupling limit is given by
the anisotropic $XXZ$ Heisenberg chain with a staggered magnetic
field
%%%%%%%%%%%%%%%%%%%%%%%%%%%%%%%%%%%%%%%%%%%%%%%%%%%%%%%%%
\begin{equation}\label{XXZ-chain}
{\cal H} =  J\sum_{n}\left( S^{x}_{n} S^{x}_{n+1} +  S^{y}_{n}
S^{y}_{n+1}+\gamma \, S^{z}_{n} S^{z}_{n+1}\right) -
h\sum_{n}\,(-1)^{n}S^{z}_{n}\, ,
\end{equation}
%%%%%%%%%%%%%%%%%%%%%%%%%%%%%%%%%%%%%%%%%%%%%%%%%%%%%%%%
where
%%%%%%%%%%%%%%%%%%%%%%%%%%%%%%%%%%%%%%%%%%%%%%%%%%%%%%%%%
\begin{equation}\label{XXZ-chain-J-gamma}
J = \frac{4t_{\uparrow} t_{\downarrow}}{U(1-x^{2})}\, , \qquad
\gamma=\frac{t^{2}_{\uparrow}+t^{2}_{\downarrow}}{2t_{\uparrow}
t_{\downarrow}}\, , \qquad h =
 \frac{2(t^{2}_{\uparrow}-t^{2}_{\downarrow})x
}{U(1-x^{2})} \, ,
\end{equation}
%%%%%%%%%%%%%%%%%%%%%%%%%%%%%%%%%%%%%%%%%%%%%%%%%%%%%%%%
and $x=\Delta/U$. For $t_{\uparrow} \neq t_{\downarrow}$ and finite
$x$, the translational symmetry is broken already at the level of
the effective spin Hamiltonian via the presence of the staggered
magnetic field. Since this represents a strongly relevant
perturbation to the spin system, the ground state is characterized
by a long-range antiferromagnetic order with explicitly broken
translational symmetry. The excitation spectrum is gapped and the
gap exhibits power-law dependence on the parameter
$h$. \cite{Luther76}

In the present paper we extend our analysis to the case of
explicitly broken translational symmetry in the on-site interaction
and derive the effective spin Hamiltonian for the one-dimensional
spin-asymmetric alternating-$U$ ionic Hubbard chain represented by
%%%%%%%%%%%%%%%%%%%%%%%%%%%%%%%%%%%%%%%%%%%%%%%%%
\begin{equation}\label{ExtHubbardHamiltonian-2}
{\cal H} =
\sum_{n,\alpha}t_{\alpha}\big(c^{\dag}_{n\alpha}
c^{\phantom{\dag}}_{n+1,\alpha} +
c^{\dag}_{n+1,\alpha}c^{\phantom{\dag}}_{n\alpha} \big)
+\,U \sum_{n}\left(\,1+(-1)^{n}\delta\,\right){\hat
\rho}^{\phantom{\dag}}_{n\uparrow} {\hat
\rho}^{\phantom{\dag}}_{n\downarrow}
+\frac{\Delta}{2}\sum_{n}\sum_{\alpha} (-1)^{n}\,{\hat
\rho}^{\phantom{\dag}}_{n\alpha} \, .
\end{equation}
%%%%%%%%%%%%%%%%%%%%%%%%%%%%%%%%%%%%%%%%%%%%%%%%
We find that up to fourth-order terms in $t_{\alpha}/U$, the
infrared behavior of the lattice fermion model
(\ref{ExtHubbardHamiltonian-2}) at half-filling is governed
by the following effective spin Hamiltonian:
%%%%%%%%%%%%%%%%%%%%%%%%%%%%%%%%%%%%%%%%%%%%%%%%%%%%%%%%%
\begin{eqnarray}\label{EffectiveHeisChainHam-1}
&{\cal H}_{eff}  = \sum\limits_{n} \Big[ J_{\perp} ( S^{x}_{n} S^{x}_{n+1}
+ S^{y}_{n} S^{y}_{n+1}) + J_{\parallel}( S^{z}_{n}
S^{z}_{n+1}-\frac{1}{4}) + J^{\prime}_{\perp}(n)\,( S^{x}_{n}
S^{x}_{n+2} + S^{y}_{n} S^{y}_{n+2})+ J^{\prime}_{\parallel}(n)\,(
S^{z}_{n} S^{z}_{n+2}-\frac{1}{4})\Big]+&\nonumber\\
%%%%%%%%%%%%%%                                                      Line 2
& +  \sum\limits_{n}\Big[ W_{\perp}(n)\left[( S^{x}_{n-1} S^{x}_{n} +
S^{y}_{n-1} S^{y}_{n}) S^{z}_{n+1}+
 S^{z}_{n-1}( S^{x}_{n} S^{x}_{n+1} +  S^{y}_{n} S^{y}_{n+1})\right] +
 W_{\parallel}(n)\, S^{z}_{n-1} S^{z}_{n} S^{z}_{n+1}\Big]-\sum\limits_{n}
 h(n)\,S^{z}_{n}\, ,&
%%%%%%%%%%%%%%%%                                                    Line 3
\end{eqnarray}
%%%%%%%%%%%%%%%%%%%%%%%%%%%%%%%%%%%%%%%%%%%%%%%%%%%%%%%%%%%%%%%%%%
where
%%%%%%%%%%%%%%%%%%%%%%%%%%%%%%%%%%%%%%%%%%%%%%%%%%
\begin{eqnarray}
   J_{\perp} &=& \frac{4t_{\uparrow} t_{\downarrow}}{U(1-\lambda^{2})}
\left[ 1
-\frac{2(t^{2}_{\uparrow}+t^{2}_{\downarrow})}{U^{2}(1-\lambda^{2})^{2}}
\left(2+ 2\lambda^{2}- \frac{1-\lambda^{2}}{1-\delta^{2}}\right)
\right],\\
%%%%%%%%%%%%%%%%%%%%%%%%%%%
J_{\parallel} &=&
\frac{2(t^{2}_{\uparrow}+t^{2}_{\downarrow})}{U(1-\lambda^{2})}-
   \frac{6(t^{4}_{\uparrow}+t^{4}_{\downarrow})}{U^{3}(1-\lambda^{2})^{3}}(1+3\lambda^{2}) -\frac{4t^{2}_{\uparrow}t^{2}_{\downarrow}}{U^{3}(1-\lambda^{2})^{3}}
   \left(5-\lambda^{2} -
   \frac{4(1-\lambda^{2})}{1-\delta^{2}}\right), \\
%%%%%%%%%%%%%%%%%%%%%%%%%%%
J^{\prime}_{\perp}(n)&=& \frac{4t^{2}_{\uparrow}
t^{2}_{\downarrow}}{U^{3}(1-\lambda^{2})^{3}} \left[2 +
2\lambda^{2}-\frac{(1-\lambda^{2})^2}{1-\delta^{2}}\right]+(-1)^{n}\frac{4\delta
t^{2}_{\uparrow}
t^{2}_{\downarrow}}{U^{3}(1-\lambda^{2})(1-\delta^{2})}\, , \\
%%%%%%%%%%%%%%%%%%%%%%%%%%%
J^{\prime}_{\parallel}(n) &=& \frac{2(t^{4}_{\uparrow} + t^{4}_{\downarrow})}{U^{3}(1-\lambda^{2})^{3}} \left[1+3\lambda^{2} + \frac{1 - \lambda^{4}}{1-\delta^{2}}\right]-
\frac{4t^{2}_{\uparrow} t^{2}_{\downarrow}}{U^{3}(1-\lambda^{2})^{2}} \,
\frac{1+\delta^{2}}{1-\delta^{2}}+ \nonumber \\
&&+\,
(-1)^{n}\frac{2\delta}{U^{3}(1-\lambda^{2})^{2}(1-\delta^{2})}\big[4t^{2}_{\uparrow}
t^{2}_{\downarrow}-(t^{4}_{\uparrow} +
t^{4}_{\downarrow})(1+\lambda^{2})\big],\\
%%%%%%%%%%%%%%%%%%%%%%%%%%%
W_{\perp}(n) &=& \frac{4\lambda \,
t_{\uparrow}t_{\downarrow}(t^2_{\uparrow} -
t^2_{\downarrow})}{U^{3}(1-\lambda^{2})^{2}} \left\{
\frac{2\delta}{1-\delta^{2}}+(-1)^{n} \left[ \frac{3+
\lambda^{2}}{1-\lambda^{2}}+ \frac{2}{1-\delta^{2}}\right] \right
\},\\
%%%%%%%%%%%%%%%%%%%%%%%%%%%
W_{\parallel}(n) &=& \frac{4\lambda (t^4_{\uparrow} -
t^4_{\downarrow})}{U^{3}(1-\lambda^{2})^{2}} \left\{
\frac{2\delta}{1-\delta^{2}}+(-1)^{n} \left[ \frac{3+
\lambda^{2}}{1-\lambda^{2}}+ \frac{2}{1-\delta^{2}}\right] \right
\},\\
%%%%%%%%%%%%%%%%%%%%%%%%%%%
h(n) &=& h_{0}+(-1)^{n}h_{1}=\frac{2\lambda\delta(t^{4}_{\uparrow} -
t^{4}_{\downarrow})}{U^{3}(1-\lambda^{2})^{2}(1-\delta^{2})}
+\nonumber\\&&+\,(-1)^{n}
\frac{2\lambda(t^{2}_{\uparrow}-t^{2}_{\downarrow})}{U(1-\lambda^{2})}\left\{1
-\frac{t^{2}_{\uparrow} +
t^{2}_{\downarrow}}{2U^{2}(1-\lambda^{2})^{2}} \left[5 \, (3+
\lambda^{2})-
   \frac{2(1-\lambda^{2})}{1-\delta^{2}}\right] \right \}, \label{Wparallel-1}
%%%%%%%%%%%%%%%%%%%%%%%%%%%
\end{eqnarray}
%%%%%%%%%%%%%%%%%%%%%%%%%%%%%%%%%%%%%%%%%%%%%%%%%%%%%%%%%%%%%%%%%%
with $\lambda=\delta+\Delta/U$.

As we observe, the leading terms ($\propto U^{-1}$) are the same as
in (\ref{XXZ-chain-J-gamma}) except that the parameter $x$ is
replaced by $\lambda$. The higher-order terms ($\propto U^{-3}$)
include the renormalization of the nearest-neighbor coupling $J$, the
next-nearest-neighbor exchange with an alternating part whose
existence is determined solely by $\delta$, and corrections to the
magnetic field alongside the less conventional three-spin terms,
all having both homogeneous and alternating parts. We also note that
the expressions of the fourth-order terms obtained in our earlier
work \cite{Grusha-Japaridze-13} are not entirely correct and they
should be replaced by the appropriate limit ($\delta=0$) of the
above results.

A detailed derivation of the expressions
(\ref{EffectiveHeisChainHam-1})-\!-(\ref{Wparallel-1}) is
presented in the following. In Sections \ref{large-U-expansion} and
\ref{HalfFill} a unitary transformation is applied to the electronic
Hamiltonian in the case of a half-filled band, eliminating hopping
processes between many-electron states with different numbers of
doubly occupied sites. In Section \ref{HubbardOperators} we briefly
discuss the Hubbard operators, which are used in the subsequent
Section \ref{SpinHamilt} to derive the effective spin Hamiltonian.
Finally, Section \ref{concl} summarizes the main results of the paper,
while the \hyperref[apndx]{Appendix} contains some technical
calculations of the spin exchange terms.

\section{The strong-coupling approach} \label{large-U-expansion}

In the strong coupling limit ($ U \gg |t| $), the perturbative treatment
of the half-filled Hubbard model based on expansion of the Hamiltonian
in powers of $t/U$ goes back to the original derivation of the
effective spin Hamiltonian to the second order by Anderson. \cite{Anderson59}
Afterwards, using different versions of the degenerate perturbation theory,
effective spin Hamiltonians up to higher orders in $t/U$ have been obtained.
In particular, Klein and Seitz \cite{Klein-Seitz-73} derived the sixth-order
spin interaction for the Hubbard chain, while Bulaevskii \cite{Bulaevskii67}
and Takahashi \cite{Takahashi77} obtained the fourth-order terms for the
half-filled Hubbard model in higher dimensions. More recently, these
perturbative methods have also been applied to Hubbard models with
more general interactions. \cite{Extended-Hubbard-LUE}

An alternative approach to construct the effective Hamiltonian is based on unitary
transformations. Harris and Lange \cite{Harris-Lange-67} used such a
transformation to obtain second-order results and to calculate spectral
properties of the Hubbard model. A transformation which systematically
incorporates higher orders in $t/U$ has been proposed by Chao, Spa\l ek,
and Ole\'s. \cite{Oles-etal-77} In their expansion, closed expressions for
the effective spin exchange are obtained to any order. However, beyond the
second order their method is not very well controlled since the
transformation of the Hamiltonian involves an approximation for the band
energies, and higher-order terms mixing different Hubbard bands
are not eliminated properly. \cite{Oles90,MacDonald90}

A consistent transformation scheme which allows one to remove all
unphysical terms and to derive the $t/U$-expansion up to any desired
order has been formulated by MacDonald, Girvin and Yoshioka. \cite{MacDonald88}
In their scheme, interaction terms which do not conserve the number
of local electron pairs are eliminated from the Hamiltonian order by
order in an iterative treatment, generating new interactions and thus
improving the accuracy of the transformation at each step. Later their
approach has been successfully employed to obtain effective spin
Hamiltonians in the case of extended versions of the Hubbard model
on a square lattice with next-nearest- and
next-next-nearest-neighbor hoppings. \cite{Tremblay04,Tremblay09}

Another consistent scheme for construction of the effective spin
Hamiltonian up to any given order in powers of $t/U$ has been developed
by Stein, \cite{Stein97} who utilized Wegner's method \cite{Wegner94}
of continuous unitary transformations with subsequent solution of the
corresponding flow equations for the half-filled Hubbard model. Later a
similar approach has been used to reveal an additional (hidden) symmetry
of the Hubbard model on any bipartite lattice. \cite{Carmelo-Ostlund-Sampaio-10}

In this paper we apply the method developed by MacDonald, Girvin and
Yoshioka for the standard Hubbard model \cite{MacDonald88} to the
Hamiltonian ${\cal H}=T+V$, where
%%%%%%%%%%%%%%%%%%%%%%%%%%%%%%%%%%%%%%%%%%%%%%%%%%%%%%%%%%%%%%%%%
\begin{eqnarray}\label{SA-Alt-U-IonHubMod}
T&=&\!\!\sum_{<\!n,m\!>}\sum_{\alpha}\, t_{\alpha}c^{\dag}_{n\alpha}c^{\phantom{\dag}}_{m\alpha} \, , \\
V & = & \frac{\Delta}{2}\,\sum_{n,\alpha}(-1)^{n}{\hat
\rho}^{\phantom{\dag}}_{n\alpha} + U_{o} \sum_{n}{\hat
\rho}^{\phantom{\dag}}_{2n+1,\uparrow} {\hat
\rho}^{\phantom{\dag}}_{2n+1,\downarrow} +U_{e} \sum_{n}{\hat
\rho}^{\phantom{\dag}}_{2n,\uparrow} {\hat
\rho}^{\phantom{\dag}}_{2n,\downarrow}\, ,
\end{eqnarray}
%%%%%%%%%%%%%%%%%%%%%%%%%%%%%%%%%%%%%%%%%%%%%%%%%%%%%%%%%%%%%%%%%
and the brackets in the sum $<\!\!n,m\!\!>$ signify that $n$ and
$m$ are labels for neighboring sites. The on-site couplings
$U_{o}=U(1-\delta)$ and $U_{e}=U(1+\delta)$ are supposed to be
strong, $U_{e} \ge U_{o} \gg |t_{\uparrow}|,|t_{\downarrow}|,\Delta$,
implying that the parameters $\delta=(U_{e}-U_{o})/(U_{e}+U_{o})$ and
$\lambda=\delta+\Delta/U$ satisfy the conditions $0 \leq  \delta < 1$,
$0 \leq  \lambda < 1$.

In the large-$U$ limit of the standard Hubbard model ($\delta=\Delta=0$)
the many-electron states are grouped according to the number of doubly
occupied sites (doublons) $N_{d}$. In the present case with $\delta,\Delta
>0$ these Hubbard subbands are split into groups of states classified
by two numbers, $N_{de}$ and $N_{do}$, representing the numbers of
doubly occupied sites on even and odd sublattices, respectively.
The hopping operator $T$ mixes the states of these subbands. The
``unmixing'' can be achieved by introducing suitable linear
combinations of the uncorrelated basis states.  The ${\cal S}$
matrix for this transformation, and the transformed Hamiltonian,
%%%%%%%%%%%%%%%%%%%%%%%%%%%%%%%%%%%%%%%%%%%%%%%%%%
\begin{equation}
{\cal H}^{\prime} = e^{{\it i}{\cal S}} {\cal H} e^{{-\it i}{\cal S}}\, ,
\end{equation}
%%%%%%%%%%%%%%%%%%%%%%%%%%%%%%%%%%%%%%%%%%%%%%%%%%
are generated by an iterative procedure, which results in an
expansion in powers of the hopping amplitudes $t_{\uparrow}$ and/or
$t_{\downarrow}$ divided by the on-site energies $U_e$ and/or $U_o$.

This expansion is based on the separation of the kinetic part of
the Hamiltonian into three terms:
%%%%%%%%%%%%%%%%%%%%%%%%%%%%%%%%%%%%%%%%%%%%%%%%%%
\begin{equation} \label{KT}
T = T_{0}+T_{1}+T_{-1} \, ,
\end{equation}
%%%%%%%%%%%%%%%%%%%%%%%%%%%%%%%%%%%%%%%%%%%%%%%%%%
where $T_{0}$ leaves the number of doubly occupied sites unchanged,
and $T_{1}$ ($T_{-1}$) increases (decreases) this number by one.
In the present case of broken translational symmetry each of these
contributions is further split into several different terms,
depending on whether the electron hops from an even to an odd
site or vice versa.

In particular, the  $T_{0}$ term is split into four separate
processes:
%%%%%%%%%%%%%%%%%%%%%%%%%%%%%%%%%%%%%%%%%%%%%%%%%%%%%%%%%%%%%%%
\begin{equation}
T_{0}=T^{pe}_{0}+T^{po}_{0}+T^{de}_{0}+T^{do}_{0}\, .
\end{equation}
%%%%%%%%%%%%%%%%%%%%%%%%%%%%%%%%%%%%%%%%%%%%%%%%%%%%%%%%%%%%%%%
Here
%%%%%%%%%%%%%%%%%%%%%%%%%%%%%%%%%%%%%%%%%%%%%%%%%%%%%%%%%%%%%%%
\begin{equation} \label{T01}
T^{pe}_{0} =  \!\!\sum_{<\!2n,m\!>}\sum_{\alpha} t_{\alpha}\,(1-{\hat
\rho}^{\phantom{\dag}}_{2n,\alphabar})\,
c^{\dag}_{2n,\alpha}c^{\phantom{\dag}}_{m\alpha}\,(1-{\hat
\rho}^{\phantom{\dag}}_{m\alphabar})
\end{equation}
%%%%%%%%%%%%%%%%%%%%%%%%%%%%%%%%%%%%%%%%%%%%%%%%%%%%%%%%%%%%%%%
and
%%%%%%%%%%%%%%%%%%%%%%%%%%%%%%%%%%%%%%%%%%%%%%%%%%%%%%%%%%%%%%%
\begin{equation} \label{T02}
T^{po}_{0} = \!\!\sum_{<\!2n+1,m\!>}\sum_{\alpha} t_{\alpha}\,(1-{\hat
\rho}^{\phantom{\dag}}_{2n+1,\alphabar})\,
c^{\dag}_{2n+1,\alpha}c^{\phantom{\dag}}_{m\alpha}\, (1-{\hat
\rho}^{\phantom{\dag}}_{m\alphabar})
\end{equation}
%%%%%%%%%%%%%%%%%%%%%%%%%%%%%%%%%%%%%%%%%%%%%%%%%%%%%%%%%%%%%%%
correspond respectively to hopping processes where an electron
with spin $\alpha$ hops from a singly occupied odd (even) site to
an empty neighboring even (odd) site, while
%%%%%%%%%%%%%%%%%%%%%%%%%%%%%%%%%%%%%%%%%%%%%%%%%%%%%%%%%%%%%%%
\begin{equation} \label{T03}
T^{de}_{0}=\!\!\sum_{<\!2n,m\!>}\sum_{\alpha}t_{\alpha}\,{\hat
\rho}^{\phantom{\dag}}_{2n,\alphabar}
c^{\dag}_{2n,\alpha}c^{\phantom{\dag}}_{m\alpha}\,{\hat
\rho}^{\phantom{\dag}}_{m\alphabar}
\end{equation}
%%%%%%%%%%%%%%%%%%%%%%%%%%%%%%%%%%%%%%%%%%%%%%%%%%%%%%%%%%%%%%%
and
%%%%%%%%%%%%%%%%%%%%%%%%%%%%%%%%%%%%%%%%%%%%%%%%%%%%%%%%%%%%%%%
\begin{equation} \label{T04}
T^{do}_{0}=\!\!\sum_{<\!2n+1,m\!>}\sum_{\alpha}t_{\alpha}\,{\hat
\rho}^{\phantom{\dag}}_{2n+1,\alphabar}\,
c^{\dag}_{2n+1,\alpha}c^{\phantom{\dag}}_{m\alpha}\, {\hat
\rho}^{\phantom{\dag}}_{m\alphabar}
\end{equation}
%%%%%%%%%%%%%%%%%%%%%%%%%%%%%%%%%%%%%%%%%%%%%%%%%%%%%%%%%%%%%%%
represent hopping processes where an electron with spin $\alpha$
hops from a doubly occupied odd (even) site to a neighboring
even (odd) site which is already occupied by another electron
with the opposite spin $\alphabar$.

In a similar fashion, the operators $T_{\pm 1}$, which change the
number of doublons by one, are also separated into even and odd
parts $T^{\phantom{0}}_{\pm 1}=T^{e}_{\pm 1}+T^{o}_{\pm 1}$, where
%%%%%%%%%%%%%%%%%%%%%%%%%%%%%%%%%%%%%%%%%%%%%%%%%%%%%%%%%%%%%%%
\begin{equation} \label{Tb1}
T^{e}_{1}=\!\!\sum_{<\!2n,m\!>}\sum_{\alpha}t_{\alpha}\, {\hat
\rho}^{\phantom{\dag}}_{2n,\alphabar}
c^{\dag}_{2n,\alpha}c^{\phantom{\dag}}_{m\alpha}\,(1-{\hat
\rho}^{\phantom{\dag}}_{m\alphabar})
\end{equation}
%%%%%%%%%%%%%%%%%%%%%%%%%%%%%%%%%%%%%%%%%%%%%%%%%%%%%%%%%%%%%%%
and
%%%%%%%%%%%%%%%%%%%%%%%%%%%%%%%%%%%%%%%%%%%%%%%%%%%%%%%%%%%%%%%
\begin{equation} \label{Ta1}
T^{o}_{1}=\!\!\sum_{<\!2n+1,m\!>}\sum_{\alpha}t_{\alpha}\, {\hat
\rho}^{\phantom{\dag}}_{2n+1,\alphabar}\,
c^{\dag}_{2n+1,\alpha}c^{\phantom{\dag}}_{m\alpha}\, (1-{\hat
\rho}^{\phantom{\dag}}_{m\alphabar})
\end{equation}
%%%%%%%%%%%%%%%%%%%%%%%%%%%%%%%%%%%%%%%%%%%%%%%%%%%%%%%%%%%%%%%
increase the number of doublons on the sublattice of even (odd)
sites, while
%%%%%%%%%%%%%%%%%%%%%%%%%%%%%%%%%%%%%%%%%%%%%%%%%%%%%%%%%%%%%%%
\begin{equation} \label{Tb-1}
T^{e}_{-1}=\!\!\sum_{<\!n,2m\!>}\sum_{\alpha}t_{\alpha}\, (1-{\hat
\rho}^{\phantom{\dag}}_{n\alphabar})\,
c^{\dag}_{n\alpha}c^{\phantom{\dag}}_{2m,\alpha}\,{\hat
\rho}^{\phantom{\dag}}_{2m,\alphabar}
\end{equation}
%%%%%%%%%%%%%%%%%%%%%%%%%%%%%%%%%%%%%%%%%%%%%%%%%%%%%%%%%%%%%%%
and
%%%%%%%%%%%%%%%%%%%%%%%%%%%%%%%%%%%%%%%%%%%%%%%%%%%%%%%%%%%%%%%
\begin{equation} \label{Ta-1}
T^{o}_{-1}=\!\!\sum_{<\!n,2m+1\!>}\sum_{\alpha}t_{\alpha}\, (1-{\hat
\rho}^{\phantom{\dag}}_{n\alphabar})\,
c^{\dag}_{n\alpha}c^{\phantom{\dag}}_{2m+1,\alpha}\,{\hat
\rho}^{\phantom{\dag}}_{2m+1,\alphabar}\, ,
\end{equation}
%%%%%%%%%%%%%%%%%%%%%%%%%%%%%%%%%%%%%%%%%%%%%%%%%%%%%%%%%%%%%%%
respectively decrease the number of doublons on the even and odd
sublattices.

One can easily check the following commutation relations:
%%%%%%%%%%%%%%%%%%%%%%%%%%%%%%%%%%%%%%%%%%%%%%%%%%%%%%%%%
\begin{equation} \label{com}
[V\, ,\,T^{s}_{\mu}]= (\mu+\delta_{\mu,0}) \Lambda_{s} T^{s}_{\mu}\, ,
\end{equation}
%%%%%%%%%%%%%%%%%%%%%%%%%%%%%%%%%%%%%%%%%%%%%%%%%%%%%%%%%
where $\mu=0,\pm 1$ and
%%%%%%%%%%%%%%%%%%%%%%%%%%%%%%%%%%%%%%%%%%%%%%%%%%%%%%%%%
\begin{equation}
\Lambda_{s}= \left\{
\begin{array}{c}
\hspace{0.9cm}\Delta,\hspace{2.7cm} s = pe \\
\hspace{0.7cm}-\Delta,\hspace{2.6cm}s = po \\
\hspace{-0.0cm}(U_{e}-U_{o})+\Delta,\hspace{1.7cm} s = de \\
\hspace{-0.1cm}-(U_{e}-U_{o})-\Delta,\hspace{1.55cm} s = do\\
\hspace{0,5cm}U_{e}+\Delta,\hspace{2.05cm} s = e \\
\hspace{0.5cm}U_{o}-\Delta,\hspace{2.05cm} s = o \end{array} \right.
\, .
\end{equation}
%%%%%%%%%%%%%%%%%%%%%%%%%%%%%%%%%%%%%%%%%%%%%%%%%%%%%%%%%
The relations (\ref{com}) reflect the fact that the energy of the
system changes by $(\mu+\delta_{\mu,0})\Lambda_{s}$ as a result of
the hopping process $T^{s}_{\mu}$.

\section{Effective Hamiltonian in the case of a half-filled band}
\label{HalfFill}

Let us now search for the unitary transformation ${\cal S}$ which
eliminates hops between states with different numbers of doubly
occupied sites in the transformed Hamiltonian
%%%%%%%%%%%%%%%%%%%%%%%%%%%%%%%%%%%%%%%%%%%%%%%%%%%%%%%%%
\begin{equation} \label{trans}
{\cal H}^{\prime}=e^{i{\cal S}} {\cal H} e^{-i{\cal S}} ={\cal H} +
[i{\cal
  S},{\cal H}]+ \frac{1}{2}[i{\cal S},[i{\cal S},{\cal H}]]+...\, .
\end{equation}
%%%%%%%%%%%%%%%%%%%%%%%%%%%%%%%%%%%%%%%%%%%%%%%%%%%%%%%%%
We follow a recursive scheme \cite{MacDonald88} which allows to
determine such a transformation to any desired order in
$t_{\alpha}/U$. The last two terms of the initial Hamiltonian
%%%%%%%%%%%%%%%%%%%%%%%%%%%%%%%%%%%%%%%%%%%%%%%%%%%%%%%%%
\begin{equation} \label{H0}
{\cal H} \equiv {\cal H}^{\prime (1)}= V + T_{0} + T_{1} + T_{-1}
\end{equation}
%%%%%%%%%%%%%%%%%%%%%%%%%%%%%%%%%%%%%%%%%%%%%%%%%%%%%%%%%
may be transformed away by choosing
%%%%%%%%%%%%%%%%%%%%%%%%%%%%%%%%%%%%%%%%%%%%%%%%%%%%%%%%%
\begin{equation} \label{S}
i{\cal S}\equiv i{\cal
S}^{(1)}=\frac{1}{U_{o}-\Delta}(T^{o}_{1}-T^{o}_{-1})+\frac{1}{U_{e}+\Delta}
(T^{e}_{1}-T^{e}_{-1}) \, .
\end{equation}
%%%%%%%%%%%%%%%%%%%%%%%%%%%%%%%%%%%%%%%%%%%%%%%%%%%%%%%%%
Substituting (\ref{H0}) and (\ref{S}) into the expansion (\ref{trans})
and applying (\ref{com}), we obtain
%%%%%%%%%%%%%%%%%%%%%%%%%%%%%%%%%%%%%%%%%%%%%%%%%%%%%%%%%
\begin{eqnarray} \label{H0-1}
{\cal H}^{\prime (2)}= V + T_{0} &+&
\frac{1}{U_{o}-\Delta}\big\{\,[T^{o}_{1},T_{0}]+[T_{0},T^{o}_{-1}]
+[T^{o}_{1},T^{o}_{-1}]\,\big\}\nonumber\\
&+&\frac{1}{U_{e}+\Delta}\left\{\,[T^{e}_{1},T_{0}]+[T_{0},T^{e}_{-1}]
+[T^{e}_{1},T^{e}_{-1}]\,\right\}\nonumber\\
&+&\frac{U_e+U_o}{2(U_e+\Delta)(U_o-\Delta)}\,\big\{\,[T^e_{1},T^{o}_{-1}]+
[T^{o}_{1},T^e_{-1}]\,\big\}\nonumber\\
&+&\frac{U_o-U_e-2\Delta}{2(U_e+\Delta)(U_o-\Delta)}\,\big\{\,[T^e_{1},T^{o}_{1}]+
[T^{o}_{-1},T^e_{-1}]\,\big\} +{\cal O}(t^{3}/U^{2}) \, .
\end{eqnarray}
%%%%%%%%%%%%%%%%%%%%%%%%%%%%%%%%%%%%%%%%%%%%%%%%%%%%%%%%%

We focus on the case of a half-filled band, where in the large-$U$
limit the lowest-energy states $\vert \Psi_{L} \rangle$ have exactly
one electron at each site. In this subspace no hops are possible without
increasing the number of doubly occupied sites. Therefore,
%%%%%%%%%%%%%%%%%%%%%%%%%%%%%%%%%%%%%%%%%%%%%%%%%%%%%%%%%
\begin{equation}\label{Elimination}
T^{e}_{-1} \vert \Psi_{L} \rangle=0\, , \hspace{1cm} T^{o}_{-1} \vert
\Psi_{L} \rangle=0\, , \hspace{1cm} T_{0} \vert \Psi_{L} \rangle=0\, ,
\end{equation}
%%%%%%%%%%%%%%%%%%%%%%%%%%%%%%%%%%%%%%%%%%%%%%%%%%%%%%%%%
and the effective Hamiltonian (\ref{H0-1}) is reduced to
%%%%%%%%%%%%%%%%%%%%%%%%%%%%%%%%%%%%%%%%%%%%%%%%%%%%%%%%%
\begin{eqnarray} \label{H2}
{\cal H}^{\prime (2)} &=& -\frac{T^{o}_{-1}T^{o}_{1}}{U_{o} -
\Delta} -\frac{T^{e}_{-1}T^{e}_{1}}{U_{e} + \Delta} + {\cal
O}(t^{3}/U^{2})\, .
\end{eqnarray}
%%%%%%%%%%%%%%%%%%%%%%%%%%%%%%%%%%%%%%%%%%%%%%%%%%%%%%%%%
To proceed further, we define:
%%%%%%%%%%%%%%%%%%%%%%%%%%%%%%%%%%%%%%%%%%%%%%%%%%%%%%%%%
\begin{equation}
T^{(k)}\left[\{s\},\{\mu\}\right]=T^{s_{1}}_{\mu_{1}}T^{s_{2}}_{\mu_{2}}
\ldots T^{s_{k}}_{\mu_{k}}\, .
\end{equation}
%%%%%%%%%%%%%%%%%%%%%%%%%%%%%%%%%%%%%%%%%%%%%%%%%%%%%%%%%
Using (\ref{com}), we can write
%%%%%%%%%%%%%%%%%%%%%%%%%%%%%%%%%%%%%%%%%%%%%%%%%%%%%%%%%
\begin{equation}
\left[\hat{V},T^{(k)}[\{s\},\{\mu\}]\right]=\sum_{i=1}^{k}\Lambda_{s_{i}}
(\mu_{i}+\delta_{\mu_{i},0})T^{(k)}[\{s\},\{\mu\}]\, .
\end{equation}
%%%%%%%%%%%%%%%%%%%%%%%%%%%%%%%%%%%%%%%%%%%%%%%%%%%%%%%%%
${\cal H}^{\prime (k)}$ contains terms of order $(t_{\alpha})^{k}$,
denoted by ${\cal H}^{\prime [k]}$, which couple states in different
subspaces. By definition $[V,{{\cal H}^{\prime}}^{[k]}]\not=0$ and
${\cal H}^{\prime [k]}$ can be expressed in the following way:
%%%%%%%%%%%%%%%%%%%%%%%%%%%%%%%%%%%%%%%%%%%%%%%%%%%%%%%%%
\begin{equation}
{\cal H}^{\prime [k]}=\sum_{\{a\}}\sum_{\{\mu\}}
C^{(k)}_{\{a\}}(\{\mu\}) T^{(k)}[\{a\},\{\mu\}]\, , \hspace{1cm}
\sum_{i=1}^{k} \mu_{i}\not=0\, .
\end{equation}
%%%%%%%%%%%%%%%%%%%%%%%%%%%%%%%%%%%%%%%%%%%%%%%%%%%%%%%%%
If at each $k$-th order step we choose ${\cal S}^{(k)}$ as
\begin{equation}
{\cal S}^{(k)}={\cal S}^{(k-1)} + {\cal S}^{[ k]} \, ,
\end{equation}
where ${\cal S}^{[ k]}$ is the solution of the equation
%%%%%%%%%%%%%%%%%%%%%%%%%%%%%%%%%%%%%%%%%%%%%%%%%%%%%%%%%
\begin{equation}
[i{\cal S}^{[ k]},V]=- {\cal H}^{\prime [k]}
\end{equation}
%%%%%%%%%%%%%%%%%%%%%%%%%%%%%%%%%%%%%%%%%%%%%%%%%%%%%%%%%
and therefore equals
%%%%%%%%%%%%%%%%%%%%%%%%%%%%%%%%%%%%%%%%%%%%%%%%%%%%%%%%%
\begin{equation}
{\cal S}^{[ k]}=-i\sum_{\{a\}}\sum_{\{\mu\}}
\frac{C^{(k)}_{\{a\}}(\{\mu\})} {\sum_{i=1}^{k} \Lambda_{a_{i}}
(\mu_{i}+\delta_{\mu_{i},0})}T^{(k)}[\{a\},\{\mu\}]\, , \hspace{1cm}
\sum_{i=1}^{k} \mu_{i}\not= 0 \, ,
\end{equation}
%%%%%%%%%%%%%%%%%%%%%%%%%%%%%%%%%%%%%%%%%%%%%%%%%%%%%%%%%
then the transformed Hamiltonian
%%%%%%%%%%%%%%%%%%%%%%%%%%%%%%%%%%%%%%%%%%%%%%%%%%%%%%%%%
\begin{equation}
{\cal H}^{\prime (k+1)}=e^{i{\cal S}^{(k)}}{\cal H}e^{-i{\cal
S}^{(k)}}
\end{equation}
%%%%%%%%%%%%%%%%%%%%%%%%%%%%%%%%%%%%%%%%%%%%%%%%%%%%%%%%%
contains terms up to the order of $(t_{\alpha})^{k}/U^{k-1}$ which
commute with the unperturbed Hamiltonian $V$ and mix states within
each subspace only.

The conditions (\ref{Elimination}) can be generalized to higher
orders
%%%%%%%%%%%%%%%%%%%%%%%%%%%%%%%%%%%%%%%%%%%%%%%%%%%%%%%%%
\begin{equation}\label{EliminationHigherOrders}
T^{(k)}[\{s\},\{\mu\}] \; \vert \Psi_{L} \rangle=0 \, ,
\end{equation}
%%%%%%%%%%%%%%%%%%%%%%%%%%%%%%%%%%%%%%%%%%%%%%%%%%%%%%%%%
if
%%%%%%%%%%%%%%%%%%%%%%%%%%%%%%%%%%%%%%%%%%%%%%%%%%%%%%%%%
\begin{equation}
M_{p}^{k}[\{\mu\}] \equiv  \sum_{i=p}^{k} \mu_{i} < 0\,
\end{equation}
%%%%%%%%%%%%%%%%%%%%%%%%%%%%%%%%%%%%%%%%%%%%%%%%%%%%%%%%%
for at least one value of $p$. Equation
(\ref{EliminationHigherOrders}) can be used to eliminate many terms
from the expansion for ${\cal H}^{\prime}$ in the subspace of
minimal $\langle V \rangle$.

The final expression of the transformed Hamiltonian ${\cal
H}^{\prime}$ up to the fourth order reads:
%%%%%%%%%%%%%%%%%%%%%%%%%%%%%%%%%%%%%%%%%%%%%%%%%%%%%%%%%
\begin{eqnarray}
\label{Effect-Ham-In-T-Oper-4} {\cal H}'^{(4)} &=&
-\frac{T^{o}_{-1}T^{o}_{1}}{U_{o} - \Delta}
-\frac{T^{e}_{-1}T^{e}_{1}}{U_{e} + \Delta}
%%%%%%%%%
-\frac{T^{o}_{-1}T^{po}_{0}T^{pe}_{0}T^{o}_{1}}{U_{o}(U_{o} - \Delta
)^{2}}-\frac{T^{o}_{-1}T^{do}_{0}T^{de}_{0}T^{o}_{1}}{U_{e}(U_{o}
-\Delta)^{2}}-\frac{T^{e}_{-1}T^{de}_{0}T^{do}_{0}T^{e}_{1}}{U_{o}(U_{e}
+ \Delta)^{2}}
\nonumber \\
\vspace{3mm}\nonumber \\
%%%%%%%%%%%%%%%%%%                                                line 1
&-& \frac{T^{e}_{-1}T^{pe}_{0}T^{po}_{0}T^{e}_{1}}{U_{e}(U_{e} +
\Delta )^{2}} -\frac{T^{o}_{-1}T^{po}_{0}T^{do}_{0}T^{e}_{1}}{(U_{o}
- \Delta )U_{o}(U_{e} + \Delta)}-
\frac{T^{e}_{-1}T^{de}_{0}T^{pe}_{0}T^{o}_{1}}{(U_{o} - \Delta
)U_{o}(U_{e} + \Delta)}\nonumber \\
\vspace{5mm}\nonumber\\
%%%%%%%%%%%%%%%%%%                                                line 2
&-&\frac{T^{o}_{-1}T^{do}_{0}T^{po}_{0}T^{e}_{1}}{(U_{o} - \Delta
)U_{e}(U_{e} + \Delta)}
-\frac{T^{e}_{-1}T^{pe}_{0}T^{de}_{0}T^{o}_{1}}{(U_{o} - \Delta
)U_{e}(U_{e} + \Delta)} -
\frac{T^{o}_{-1}T^{e}_{-1}T^{o}_{1}T^{e}_{1}}{(U_{e} + \Delta)(U_{o}
- \Delta )(U_{e}+U_{o})}
\nonumber \\
\vspace{3mm}\nonumber \\
%%%%%%%%%%%%%%%%%%                                                line 3
&-&\frac{T^{e}_{-1}T^{o}_{-1}T^{e}_{1}T^{o}_{1}}{(U_{e} +
\Delta)(U_{o} - \Delta
)(U_{e}+U_{o})}-\frac{T^{e}_{-1}T^{o}_{-1}T^{o}_{1}T^{e}_{1}}{(U_{e}
+ \Delta)^{2}(U_{e} + U_{o})} -
\frac{T^{o}_{-1}T^{e}_{-1}T^{e}_{1}T^{o}_{1}}{(U_{o} -
\Delta)^{2}(U_{e} + U_{o})}
\nonumber\\
\vspace{3mm}\nonumber \\
%%%%%%%%%                                                          line 4
&-&\frac{T^{o}_{-1}T^{o}_{-1}T^{o}_{1}T^{o}_{1}}{2(U_{o} - \Delta
)^{3}} -\frac{T^{e}_{-1}T^{e}_{-1}T^{e}_{1}T^{e}_{1}}{2(U_{e} +
\Delta )^{3}} +\frac{T^{o}_{-1}T^{o}_{1}T^{o}_{-1}T^{o}_{1}}{(U_{o}
- \Delta )^{3}}+\frac{T^{e}_{-1}T^{e}_{1}T^{e}_{-1}T^{e}_{1}}{(U_{e}
+ \Delta )^{3}}
\nonumber \\
\vspace{3mm}\nonumber \\
%%%%%%%%%%%%%%%%%%                                                line 5
&+&\frac{U_{o}+U_{e}}{2(U_{o} - \Delta )^{2}(U_{e} + \Delta)^{2}}
\left[T^{o}_{-1}T^{o}_{1}T^{e}_{-1}T^{e}_{1}+T^{e}_{-1}T^{e}_{1}T^{o}_{-1}T^{o}_{1}\right]
\, .
\end{eqnarray}
%%%%%%%%%%%%%%%%%%%%%%%%%%%%%%%%%%%%%%%%%%%%%%%%%%%%%%%%%

\section{Hubbard operators} \label{HubbardOperators}

To handle the effects of strong interaction properly, it is
important to know whether at the beginning or at the end of a given
hopping process a particular site is doubly occupied or not. For
this purpose one introduces the so-called Hubbard operators
\cite{Hubbard-Operators-Book} $X_{n}^{ab}$, which are defined at
each site of the lattice and describe all possible transitions
between the local basis states $\vert a \rangle$,\,$\vert b
\rangle$: unoccupied $\vert 0 \rangle$, singly occupied with an
up-spin $\vert \!\! \uparrow \rangle$ or a down-spin $\vert \!\!
\downarrow \rangle$ electron, and doubly occupied $\vert 2 \rangle$.
The original electron creation (annihilation) operators can be
expressed in terms of Hubbard operators in the following way:
%%%%%%%%%%%%%%%%%%%%%%%%%%%%%%%%%%%%%%%%%%%%%%%%%%%%%%%%%%%%%%%%%
\begin{equation}
c^{\dagger}_{n\alpha}=X^{\alpha 0}_{n} + \eta(\alpha)\, X^{2 \alphabar}_{n} \, ,
\hspace{1cm} c_{n\alpha}=X^{0\alpha}_{n} + \eta(\alpha)\,
X^{\alphabar 2}_{n}\, ,
\end{equation}
%%%%%%%%%%%%%%%%%%%%%%%%%%%%%%%%%%%%%%%%%%%%%%%%%%%%%%%%%%%%%%%%%
where $\eta(\alpha)=\left \{ \begin{array}{cc} 1 & \mbox{if }
\alpha=\uparrow, \\ -1 & \mbox{if } \alpha=\downarrow. \end{array} \right.$

Conversely, in terms of creation (annihilation) operators the
Hubbard operators have the form:
%%%%%%%%%%%%%%%%%%%%%%%%%%%%%%%%%%%%%%%%%%%%%%%%%%%%%%%%%%%%%%%%%
\begin{equation}
\begin{array}{ll}
X_{n}^{\alpha 0} = c^{\dag}_{n \alpha}(1-{\hat
\rho}^{\phantom{0}}_{i \alphabar})\, , & X_{n}^{2\alpha}=\eta(\alphabar)\,
c^{\dag}_{i \alphabar}{\hat \rho}^{\phantom{\dag}}_{i \alpha}\, , \\
X_{n}^{\alpha \alphabar} =  c^{\dag}_{n \alpha} c^{\phantom{0}}_{n \alphabar}\, ,
& X_{n}^{20} =\eta(\alphabar)\, c^{\dag}_{n \alphabar}c^{\dag}_{n \alpha}\, , \\
X_{n}^{00} = (1-{\hat \rho}^{\phantom{\dag}}_{n \uparrow}) (1-{\hat
\rho}^{\phantom{\dag}}_{n \downarrow}), \qquad & X_{n}^{22} = {\hat
\rho}^{\phantom{\dag}}_{n \uparrow}\,{\hat
\rho}^{\phantom{\dag}}_{n \downarrow} \, ,  \\
X_{n}^{\alpha\alpha} =  {\hat \rho}^{\phantom{\dag}}_{n \alpha} (1-{\hat
\rho}^{\phantom{\dag}}_{n \alphabar})\, . &
\end{array}
\end{equation}
%%%%%%%%%%%%%%%%%%%%%%%%%%%%%%%%%%%%%%%%%%%%%%%%%%%%%%%%%%%%%%%%%
%%%%%%%%%%%%%%%%%%%%%%%%%%%%%%%%%%%%%%%%%%%%%%%%%%%%%%%%%%%%%%%%%
The Hubbard operators containing an even (odd) number of electron
creation/annihilation operators are Bose-like (Fermi-like)
operators. They obey the following on-site multiplication rule
%%%%%%%%%%%%%%%%%%%%%%%%%%%%%%%%%%%%%%%%%%%%%%%%%%%%%%%%%%%%%%%%%
\begin{equation}
X^{pq}_{n}X^{rs}_{n}=\delta_{qr}X^{ps}_{n}
\end{equation}
%%%%%%%%%%%%%%%%%%%%%%%%%%%%%%%%%%%%%%%%%%%%%%%%%%%%%%%%%%%%%%%%%
and commutation relations
%%%%%%%%%%%%%%%%%%%%%%%%%%%%%%%%%%%%%%%%%%%%%%%%%%%%%%%%%%%%%%%%%
\begin{equation}
[X^{pq}_{n},X^{rs}_{m}]_{\pm}=\delta_{nm}(\delta_{qr}X^{ps}_{m} \pm
\delta_{ps}X^{rq}_{m})\, ,
\end{equation}
%%%%%%%%%%%%%%%%%%%%%%%%%%%%%%%%%%%%%%%%%%%%%%%%%%%%%%%%%%%%%%%%%
where the upper sign stands for the case when both operators are
Fermi-like, otherwise the lower sign should be adopted.

It is straightforward to represent the hopping terms introduced in
Section \ref{large-U-expansion} by the Hubbard operators:
%%%%%%%%%%%%%%%%%%%%%%%%%%%%%%%%%%%%%%%%%%%%%%%%%%%%%%%%%%%%%%%%%
\begin{equation}\label{T00withX}
\begin{array}{ll}
T^{po}_{0}=\sum\limits_{<\!2n+1,m\!>}\sum\limits_{\alpha} t_{\alpha}\,
X_{2n+1}^{\alpha 0}X_{m}^{0\alpha} \, , &
T^{pe}_{0}=\sum\limits_{<\!2n,m\!>}\sum\limits_{\alpha}
t_{\alpha}\, X_{2n}^{\alpha 0}X_{m}^{0\alpha}\, , \\
%%%%%
T^{do}_{0}=\sum\limits_{<\!2n+1,m\!>}\sum\limits_{\alpha} t_{\alpha}\,
X_{2n+1}^{2 \alphabar}X_{m}^{\alphabar 2}\, , &
T^{de}_{0}=\sum\limits_{<\!2n,m\!>}\sum\limits_{\alpha} t_{\alpha}\,
X_{2n}^{2 \alphabar} X_{m}^{\alphabar 2}\, , \\
%%%%%
T^{o}_{1}=\sum\limits_{<\!2n+1,m\!>}\sum\limits_{\alpha}\eta(\alpha)\, t_{\alpha}\,
X_{2n+1}^{2 \alphabar}X_{m}^{0\alpha}\, , &
T^{e}_{1}=\sum\limits_{<\!2n,m\!>}\sum\limits_{\alpha}\eta(\alpha)\, t_{\alpha}\,
X_{2n}^{2 \alphabar}X_{m}^{0 \alpha}\, , \\
%%%%%
T^{o}_{-1}=\sum\limits_{<\!n,2m+1\!>}\sum\limits_{\alpha}\eta(\alpha)\, t_{\alpha}\,
X_{n}^{\alpha 0}X_{2m+1}^{\alphabar 2}\, , \qquad &
T^{e}_{-1}=\sum\limits_{<\!n,2m\!>}\sum\limits_{\alpha} \eta(\alpha)\, t_{\alpha}\,
X_{n}^{\alpha 0}X_{2m}^{\alphabar 2}\, .
\end{array}
\end{equation}
%%%%%%%%%%%%%%%%%%%%%%%%%%%%%%%%%%%%%%%%%%%%%%%%%%%%%%%%%%%%%%%%%
One also easily verifies that the $X$-operators describing the
transitions between singly occupied states can be rewritten in terms
of spin $S=1/2$ operators as
%%%%%%%%%%%%%%%%%%%%%%%%%%%%%%%%%%%%%%%%%%%%%%%%%%%%%%%%%%%%%%%%%
\begin{equation} \label{SpinXoperators}
\begin{array}{ll}
X^{\uparrow\downarrow}_{n} = c^{\dagger}_{n \uparrow}
c^{\phantom{\dagger}}_{n \downarrow}= S^{+}_{n}\, , \qquad &
X^{\downarrow\uparrow}_{n}= c^{\dagger}_{n \downarrow}
c^{\phantom{\dagger}}_{n \uparrow} = S^{-}_{n}\, , \\
X^{\uparrow\uparrow}_{n} = \frac{1}{2}+ S^{z}_{n} \, , &
X^{\downarrow\downarrow}_{n} = \frac{1}{2} - S^{z}_{n}\, .
\end{array}
\end{equation}
%%%%%%%%%%%%%%%%%%%%%%%%%%%%%%%%%%%%%%%%%%%%%%%%%%%%%%%%%%%%%%%%

\section{The Spin Hamiltonian} \label{SpinHamilt}

Using the relations (\ref{T00withX})-\!-(\ref{SpinXoperators}),
it is straightforward to rewrite the products of $T$-terms in
(\ref{Effect-Ham-In-T-Oper-4}) via the Hubbard and hence the spin
$S=1/2$ operators. We first consider the simplest two-component
$T$-terms at great length to elucidate the procedure for more
complicated contributions.

\subsection{The second-order terms} \label{2ndOrder}

Let us start from the hopping term which corresponds to creation and
subsequent annihilation of a single doublon on an even $2n$-th
site. Since the electron hopping is restricted to nearest neighbor
sites, this process only includes electrons located on two
neighboring sites $ 2n \pm 1 $, and is given by:
%%%%%%%%%%%%%%%%%%%%%%%%%%%%%%%%%%%%%
\begin{eqnarray}
T^{e}_{-1}T^{e}_{1}&=& \sum_{n=1}^{L/2}\sum_{q = \pm 1}\sum_{\alpha}
\Big[\,t^{2}_{\alpha}\, X_{2n+q}^{\alpha 0}X_{2n}^{\alphabar 2}
X_{2n}^{2 \alphabar} X_{2n+q}^{0\alpha} -
t_{\alpha}t_{\alphabar}\,X_{2n+q}^{\alpha 0}X_{2n}^{\alphabar
2}X_{2n}^{2\alpha}X_{2n+q}^{0 \alphabar}\,\Big]\nonumber\\
%%%%%%%%%%%%%                                                     line 1
&=& \sum_{n=1}^{L/2}\sum_{q=\pm 1}\sum_{\alpha}
\Big[\,t^{2}_{\alpha}\, X_{2n+q}^{\alpha \alpha}X_{2n}^{\alphabar
\alphabar} -t_{\alpha}t_{\alphabar}\,X_{2n+q}^{\alpha \alphabar}
X_{2n}^{\alphabar \alpha} \Big]\nonumber\\
%%%%%%%%%%%%%                                                     line 2
&=& \sum_{n=1}^{L/2}\sum_{q=\pm 1} \Big[\,t^{2}_{\uparrow}\,
X_{2n+q}^{\uparrow\uparrow}X_{2n}^{\downarrow\downarrow}
+ t^{2}_{\downarrow} \,
X_{2n+q}^{\downarrow\downarrow}X_{2n}^{\uparrow\uparrow}
-t_{\uparrow}t_{\downarrow}\left(X_{2n+q}^{\uparrow\downarrow}
X_{2n}^{\downarrow\uparrow} + X_{2n+q}^{\downarrow\uparrow}
X_{2n}^{\uparrow\downarrow} \right)\big] \,\nonumber\\
%%%%%%%%%%%%%                                                     line 3
&=& \sum_{n=1}^{L/2}\sum_{q=\pm 1}\Big[\,t^{2}_{\uparrow}\,
\big(\frac{1}{2}+ S^{z}_{2n+q}\big)
  \big(\frac{1}{2}-S^{z}_{2n}\big) +
  t^{2}_{\downarrow}\,\big(\frac{1}{2}- S^{z}_{2n+q}\big)
  \big(\frac{1}{2}+S^{z}_{2n}\big)\,\nonumber\\
&&\hspace{15mm}
-t_{\uparrow}t_{\downarrow}\left(S^{+}_{2n+q}S^{-}_{2n} +
 S^{-}_{2n+q}S^{+}_{2n}\right) \Big] \,\nonumber\\
%%%%%%%%%%%%%%%%%%%%%%%%%%%%%%%%%%%%%%%%
&=& \sum_{n=1}^{L/2}\sum_{q=\pm 1}\Big[ \left(t^{2}_{\uparrow} +
t^{2}_{\downarrow}\right)\big(\frac{1}{4}-S^{z}_{2n}S^{z}_{2n+q}\big)
- \frac{1}{2}\left(t^{2}_{\uparrow} - t^{2}_{\downarrow}\right)\big(
 S^{z}_{2n}-S^{z}_{2n+q}\big)\,\nonumber\\
&&\hspace{15mm}  -2t_{\uparrow}t_{\downarrow}(S^{x}_{2n}S^{x}_{2n+q}
  + S^{y}_{2n}S^{y}_{2n+q}) \Big]\nonumber\\
%%%%%%%%%%%%%%%%%%%%%%%%%%%%%%%%%%%%%%%%
&=& \sum_{n=1}^{L}
\Big[-2t_{\uparrow}t_{\downarrow}\left(S^{x}_{n}S^{x}_{n+1}
  + S^{y}_{n}S^{y}_{n+1}\right)-\left(t^{2}_{\uparrow} +
t^{2}_{\downarrow}\right)\big(S^{z}_{n}S^{z}_{n+1}-\frac{1}{4}\big)
 - (-1)^{n}\left(t^{2}_{\uparrow} -
t^{2}_{\downarrow}\right)S^{z}_{n} \Big]\, .
\end{eqnarray}
%%%%%%%%%%%%%%%%%%%%%%%%%%%%%%%%%%%%%%%%

As to the second term $ T^{o}_{-1}T^{o}_{1} $, which describes
creation and annihilation of a pair on an odd site, the calculation
is essentially the same, with the only difference being that one
needs to make a replacement $ 2n \leftrightarrow 2n+1 $. Below this
shift is absorbed in $q$:
%%%%%%%%%%%%%%%%%%%%%%%%%%%%%%%%%%%%%
\begin{eqnarray}
T^{o}_{-1}T^{o}_{1}&=& \sum_{n=1}^{L/2}\sum_{q = 0,2}\sum_{\alpha}
\Big[\,t^{2}_{\alpha}\, X_{2n+q}^{\alpha 0}X_{2n+1}^{\alphabar 2}
X_{2n+1}^{2 \alphabar} X_{2n+q}^{0 \alpha} -
t_{\alpha}t_{\alphabar}\,X_{2n+q}^{\alpha 0}X_{2n+1}^{\alphabar
2}X_{2n+1}^{2\alpha}X_{2n+q}^{0 \alphabar}\,\Big]\nonumber\\
%%%%%%%%%%%%%                                                    line 1
&=& \sum_{n=1}^{L/2}\sum_{q=0,2}\Big[\,t^{2}_{\uparrow}\,
\big(\frac{1}{2}+ S^{z}_{2n+q}\big)
  \big(\frac{1}{2}-S^{z}_{2n+1}\big) +
  t^{2}_{\downarrow}\,\big(\frac{1}{2}- S^{z}_{2n+q}\big)
  \big(\frac{1}{2}+S^{z}_{2n+1}\big)\,\nonumber\\
&&\hspace{15mm}
-t_{\uparrow}t_{\downarrow}\left(S^{+}_{2n+q}S^{-}_{2n+1} +
 S^{-}_{2n+q}S^{+}_{2n+1}\right) \Big] \,\nonumber\\
 %%%%%%%%%%%%%                                                    line 2
 &=& \sum_{n=1}^{L/2}\sum_{q=0,2}\Big[ \left(t^{2}_{\uparrow} +
t^{2}_{\downarrow}\right)\big(\frac{1}{4}-S^{z}_{2n+q}S^{z}_{2n+1}\big)
+ \frac{1}{2}\left(t^{2}_{\uparrow} - t^{2}_{\downarrow}\right)\big(
 S^{z}_{2n+q}-S^{z}_{2n+1}\big)\,\nonumber\\
&&\hspace{15mm}
-2t_{\uparrow}t_{\downarrow}(S^{x}_{2n+q}S^{x}_{2n+1}
  + S^{y}_{2n+q}S^{y}_{2n+1}) \Big]\nonumber\\
  &=& \sum_{n=1}^{L}
\Big[-2t_{\uparrow}t_{\downarrow}(S^{x}_{n}S^{x}_{n+1}
  + S^{y}_{n}S^{y}_{n+1})-\left(t^{2}_{\uparrow} +
t^{2}_{\downarrow}\right)\big(S^{z}_{n}S^{z}_{n+1}-\frac{1}{4}\big)
+ (-1)^{n}\left(t^{2}_{\uparrow} -
t^{2}_{\downarrow}\right)S^{z}_{n} \Big]\, .
  %%%%%%%%%%%%%                                                    line 3
\end{eqnarray}
%%%%%%%%%%%%%%%%%%%%%%%%%%%%%%%%%%%%%%%%

The combination of these two processes yields the second-order
effective spin Hamiltonian:
%%%%%%%%%%%%%%%%%%%%%%%%%%%%%%%%%%%
%%%%%%%%%%%%%%%%%%%%%%%%%%%%%%%%%%%
\begin{eqnarray}\label{EffHam2ndOrder}
{\cal H}_{eff}^{(2)} &=& -\frac{1}{U_{o}-\Delta}T^{o}_{-1}T^{o}_{1}
-\frac{1}{U_{e}+\Delta}T^{e}_{-1}T^{e}_{1}=
\nonumber \\
&=&J \sum_{n}\left[ S^{x}_{n} S^{x}_{n+1} +  S^{y}_{n} S^{y}_{n+1} +
\gamma \left(S^{z}_{n} S^{z}_{n+1}-1/4\right)\right] -
h\sum_{n}(-1)^{n}S^{z}_{n}\, ,
\end{eqnarray}
%%%%%%%%%%%%%%%%%%%%%%%%%%%%%%%%%%%%%%%%%%
where
%%%%%%%%%
%%%%%%%%%%%%%%%%%%%%%%%%%%%%%%%%%%%%%%%%%%%%%%%%%%%%%%%%%
\begin{equation}\label{J-gamma-h}
J = \frac{4 t_{\uparrow} t_{\downarrow}}{U(1-\lambda^{2})}; \quad
\gamma = \frac{t^{2}_{\uparrow} +t^{2}_{\downarrow}}{2 t_{\uparrow}
t_{\downarrow}}; \quad h = \frac{2\lambda(t^{2}_{\uparrow} -
t^{2}_{\downarrow})}{U(1-\lambda^{2})}\, .
\end{equation}
%%%%%%%%%%%%%%%%%%%%%%%%%%%%%%%%%%%%%%%%%%%%%%%%%%%%%%%%
As we see, the second-order effective Hamiltonian, which describes
the spin degrees of freedom of the initial lattice fermion model, is
the Hamiltonian of spin $S=1/2$ frustrated $XXZ$ Heisenberg chain in
the presence of a staggered magnetic field. The amplitude of this
field is proportional to the product of the parameter $\lambda$
quantifying the broken translational symmetry of the underlying
fermion model, and the spin-dependent hopping asymmetry parameter
$t_{\uparrow} - t_{\downarrow}$. Thus, in contrast with the
spin-isotropic case ($t_{\uparrow} = t_{\downarrow}$), the infrared
properties of the spin-asymmetric model are described by a
Hamiltonian with an explicitly broken translational symmetry.

It is instructive to check several limiting cases. In the case of
spin-symmetric electron hopping ($t_{\uparrow} = t_{\downarrow}=t$),
the effective Hamiltonian (\ref{EffHam2ndOrder}) reduces to the
Hamiltonian of the isotropic ($SU(2)$-invariant) Heisenberg chain
%%%%%%%%%%%%%%%%%%%%%%%%%%%%%%%%%%%
\begin{equation}\label{HeisenbergHamiltonian}
{\cal H}_{eff}^{(2)} = J \sum_{n} {\bf S}_{n}\cdot {\bf S}_{n+1}\, ,
\end{equation}
%%%%%%%%%%%%%%%%%%%%%%%%%%%%%%%%%%%%%%%%%%
with a uniform exchange constant $ J=4t^{2}/U(1-\lambda^{2})$.
Thus, even if the translational symmetry of the underlying fermion
model is broken ($ \lambda \neq 0 $), the second-order effective
spin Hamiltonian remains translationally invariant.

In the complementary case of the Hubbard model with spin-dependent
hopping ($\lambda = 0,\, t_{\uparrow} \neq t_{\downarrow}$), the
second-order effective Hamiltonian properly reflects the broken spin
symmetry and is given by the Hamiltonian of anisotropic
($U(1)$-invariant) Heisenberg chain
%%%%%%%%%%%%%%%%%%%%%%%%%%%%%%%%%%%
\begin{equation}\label{XYZ-HeisenbergHamiltonian}
{\cal H}_{eff}^{(2)} = J \sum_{n}\left( S^{x}_{n} S^{x}_{n+1} +
S^{y}_{n} S^{y}_{n+1} + \gamma \, S^{z}_{n} S^{z}_{n+1}\right) \, ,
\end{equation}
%%%%%%%%%%%%%%%%%%%%%%%%%%%%%%%%%%%%%%%%%%%%%%%%%%%%%%%%%
with the anistropy parameter $\gamma = (t^{2}_{\uparrow}
+t^{2}_{\downarrow})/2 t_{\uparrow} t_{\downarrow}>1$. \cite{Fath95}

Finally, in the limiting case of the Falicov-Kimball model
($t_{\downarrow}=0$), the second-order effective spin Hamiltonian
reduces to the Ising model in a staggered magnetic field:
%%%%%%%%%%%%%%%%%%%%%%%%%%%%%%%%%%%
\begin{equation}
{\cal H}_{eff}^{(2)} =  \sum_{n}  \left(\, J_{\parallel}\,S^{z}_{n}
S^{z}_{n+1} - (-1)^{n}\,h\,S^{z}_{n}\,\right)
,\label{EffHam-2-FK-Ising}
\end{equation}
%%%%%%%%%%%%%%%%%%%%%%%%%%%%%%%%%%%%%%%%%%
where
%%%%%%%%%
%%%%%%%%%%%%%%%%%%%%%%%%%%%%%%%%%%%%%%%%%%%%%%%%%%%%%%%%%
\begin{equation}\label{Jparallel-h}
J_{\parallel} = \frac{2t^{2}_{\uparrow}}{U(1-\lambda^{2})} ; \quad h
= \frac{2\lambda t^{2}_{\uparrow}}{U(1-\lambda^{2})}\, .
\end{equation}
%%%%%%%%%%%%%%%%%%%%%%%%%%%%%%%%%%%%%%%%%%%%%%%%%%%%%%%%

The physical mechanism responsible for appearance of the staggered
magnetic field in the effective spin Hamiltonian
(\ref{EffHam2ndOrder}) can easily be traced in the ultimate
limit of the Falicov-Kimball model, however the argument remains
valid also for arbitrary $t_{\uparrow} > t_{\downarrow}>0$. Due to
the doubling of the lattice unit cell, energetically it is
preferable to locate all immobile fermions on odd sites, while the
mobile up-spin fermions will predominantly occupy even sites. In
this limit, the process of creation and annihilation of a doublon
takes place only on odd sites and gives rise to the following
Ising-type spin exchange parameter
$J_{\parallel}^{(1)}=2t_{\uparrow}^{2}/(U_{o}-\Delta)$, while in the
opposite case, where all immobile spins are located on even sites,
the same process yields the exchange constant
$J_{\parallel}^{(2)}=2t_{\uparrow}^{2}/(U_{e}+\Delta)$. The
difference between the exchange energies for these two patterns
equals
%%%%%%%%%%%%%%%%%%%%%%%%%%%%%%%%%%%%%%%%%%%%%%%%%%%%%%%%%
\begin{equation}\label{DeltaJ}
J_{\parallel}^{(1)}-J_{\parallel}^{(2)}=\frac{4\lambda
t^{2}_{\uparrow}}{U(1-\lambda^{2})}=2h.
\end{equation}
%%%%%%%%%%%%%%%%%%%%%%%%%%%%%%%%%%%%%%%%%%%%%%%%%%%%%%%%

\subsection{The fourth-order terms}

The same technique as the one employed in the previous section can
be used to rewrite the products of four $T$-terms in the effective
Hamiltonian (\ref{Effect-Ham-In-T-Oper-4}) via the spin $S=1/2$
operators. There are 18 terms of this type. It is convenient to
unite them in groups characterized by the similarity of the hopping
processes and by the number of created doublons at the intermediate
steps.

\subsubsection{Group A: Four-{\it T} product terms of \texorpdfstring{the form
$T_{-1}T_{0}T_{0}T_{1}$}{type 1}}

There are eight terms of this type in the  effective Hamiltonian
(\ref{Effect-Ham-In-T-Oper-4}). In these processes the number of
created doubly occupied sites is one. Four terms correspond to
processes where the doublon is created and eventually annihilated on
the same site, while the other four terms describe processes where
the doublon is created on an odd (even) site and annihilated on the
neighboring even (odd) site. The  calculations are straightforward
and one obtains the following expressions for the operators (the
details can be found in the \hyperref[apndx]{Appendix}):
%%%%%%%%%%%%%%%%%%%%%%%%%%%%%%%%%%%%%
\begin{eqnarray}\label{4T-A1}
&&  \quad  T^{o}_{-1}T^{po}_{0}T^{pe}_{0}T^{o}_{1}
    = \sum_{n} \Big[\, -t_{\uparrow}t_{\downarrow}(t^2_{\uparrow}+t^2_{\downarrow})
    (S^{x}_{n}S^{x}_{n+1} + S^{y}_{n}S^{y}_{n+1}) -  2 t^2_{\uparrow}t^2_{\downarrow}
\big( S^{z}_{n}S^{z}_{n+1} - 1/4 \big)
+\frac{t^4_{\uparrow}-t^4_{\downarrow}}{2}(-1)^n S^{z}_{n} -
\nonumber\\
   &&  \hspace{28mm} -   (t^2_{\uparrow}-t^2_{\downarrow})^2
\big( S^{z}_{2n-1}S^{z}_{2n+1} - 1/4 \big)
-2(t^4_{\uparrow}-t^4_{\downarrow}) \big( S^{z}_{2n-1}S^{z}_{2n} -
1/4 \big)S^{z}_{2n+1}  - \nonumber\\
   &&\hspace{28mm}  - 2t_{\uparrow}t_{\downarrow}(t^2_{\uparrow}-t^2_{\downarrow})
   [(S^{x}_{2n-1}S^{x}_{2n} + S^{y}_{2n-1}S^{y}_{2n})S^{z}_{2n+1}+
   S^{z}_{2n-1}(S^{x}_{2n}S^{x}_{2n+1} +
   S^{y}_{2n}S^{y}_{2n+1})]\,\Big]\, ,\\
\nonumber\\
%%%%%%%%%%%%%%%%%%%%%%%%%%%%%%%%%%%%%
%%%%%%%%%%%%%%%%%%%%%%%%%%%%%%%%%%%%%
&& \quad T^{o}_{-1}T^{do}_{0}T^{de}_{0}T^{o}_{1} = \sum_{n}\Big[\, -
t_{\uparrow}t_{\downarrow}(t^2_{\uparrow}+t^2_{\downarrow})(S^{x}_{n}S^{x}_{n+1}
+ S^{y}_{n}S^{y}_{n+1}) -  2 t^2_{\uparrow}t^2_{\downarrow} \big(
S^{z}_{n}S^{z}_{n+1} - 1/4 \big) +
\frac{t^4_{\uparrow}-t^4_{\downarrow}}{2}(-1)^{n} S^{z}_{n} -\nonumber\\
    && \hspace{28mm} -   (t^2_{\uparrow}-t^2_{\downarrow})^2
\big( S^{z}_{2n}S^{z}_{2n+2} - 1/4 \big)   +
2(t^4_{\uparrow}-t^4_{\downarrow}) \big( S^{z}_{2n}S^{z}_{2n+1} -
1/4 \big)S^{z}_{2n+2}+
    \nonumber\\
&& \hspace{28mm}+
2t_{\uparrow}t_{\downarrow}(t^2_{\uparrow}-t^2_{\downarrow})[(S^{x}_{2n}S^{x}_{2n+1}
+ S^{y}_{2n}S^{y}_{2n+1})S^{z}_{2n+2} +
S^{z}_{2n}(S^{x}_{2n+1}S^{x}_{2n+2} + S^{y}_{2n+1}S^{y}_{2n+2})] \,
\Big]\, ,\label{4T-A2}\\
\nonumber\\
%%%%%%%%%%%%%%%%%%%%%%%%%%%%%%%%%%%%%%
%%%%%%%%%%%%%%%%%%%%%%%%%%%%%%%%%%%%%%
&&  \quad T^{e}_{-1}T^{pe}_{0}T^{po}_{0}T^{e}_{1} =
    \sum_{n}\Big[\, - t_{\uparrow}t_{\downarrow}(t^2_{\uparrow}+t^2_{\downarrow})
    (S^{x}_{n}S^{x}_{n+1} + S^{y}_{n}S^{y}_{n+1}) - 2 t^2_{\uparrow}t^2_{\downarrow}
\big( S^{z}_{n}S^{z}_{n+1} - 1/4 \big)-
\frac{t^4_{\uparrow}-t^4_{\downarrow}}{2}(-1)^{n} S^{z}_{n} - \nonumber\\
&& \hspace{28mm}  -  (t^2_{\uparrow}-t^2_{\downarrow})^2 \big(
S^{z}_{2n}S^{z}_{2n+2} - 1/4 \big)  -
2(t^4_{\uparrow}-t^4_{\downarrow}) \big( S^{z}_{2n}S^{z}_{2n+1} -1/4 \big)S^{z}_{2n+2}- \nonumber\\
   && \hspace{28mm}- 2t_{\uparrow}t_{\downarrow}(t^2_{\uparrow}-t^2_{\downarrow})
   [(S^{x}_{2n}S^{x}_{2n+1} + S^{y}_{2n}S^{y}_{2n+1})S^{z}_{2n+2} +
   S^{z}_{2n}(S^{x}_{2n+1}S^{x}_{2n+2} +
   S^{y}_{2n+1}S^{y}_{2n+2})]\, \Big]\, ,\label{4T-A3}  \\
%%%%%%%%%%%%%%%%%%%%%%%%%%%%%%%%%%%%%%%%%%%
%%%%%%%%%%%%%%%%%%%%%%%%%%%%%%%%%%%%%%%%%%%
   && \quad   T^{e}_{-1}T^{de}_{0}T^{do}_{0}T^{e}_{1} =
\sum_{n}\Big[\,-
t_{\uparrow}t_{\downarrow}(t^2_{\uparrow}+t^2_{\downarrow})(S^{x}_{n}S^{x}_{n+1}
+ S^{y}_{n}S^{y}_{n+1}) - 2 t^2_{\uparrow}t^2_{\downarrow} \big(
S^{z}_{n}S^{z}_{n+1} - 1/4 \big) -
\frac{t^4_{\uparrow}-t^4_{\downarrow}}{2}(-1)^{n} S^{z}_{n}-\nonumber\\
    &&\hspace{28mm}  -  (t^2_{\uparrow}-t^2_{\downarrow})^2
\big( S^{z}_{2n-1}S^{z}_{2n+1} - 1/4 \big)
+2(t^4_{\uparrow}-t^4_{\downarrow}) \big( S^{z}_{2n-1}S^{z}_{2n} -
1/4 \big)S^{z}_{2n+1}+
    \nonumber\\
&&\hspace{28mm}   +
2t_{\uparrow}t_{\downarrow}(t^2_{\uparrow}-t^2_{\downarrow})[(S^{x}_{2n-1}S^{x}_{2n}
+ S^{y}_{2n-1}S^{y}_{2n})S^{z}_{2n+1} +
S^{z}_{2n-1}(S^{x}_{2n}S^{x}_{2n+1} + S^{y}_{2n}S^{y}_{2n+1})] \,
\Big]\, ,\label{4T-A4}  \\
%%%%%%%%%%%%%%%%%%%%%%%%%%%%%
%%%%%%%%%%%%%%%%%%%%%%%%%%%%%
&& \quad    T^{e}_{-1}T^{de}_{0}T^{pe}_{0}T^{o}_{1}
   = \sum_{n}\Big[\,-
   t_{\uparrow}t_{\downarrow}(t^2_{\uparrow}+t^2_{\downarrow})
   \big(S_{n}^{x}S_{n+1}^{x}+ S_{n}^{y}S_{n+1}^{y}\big) -
   2 t^2_{\uparrow}t^2_{\downarrow}
    \left( S^{z}_{n}S^{z}_{n+1} - 1/4 \right) +  \nonumber\\
   && \hspace{28mm}+  i\, t_{\uparrow}t_{\downarrow}(t^2_{\uparrow}-t^2_{\downarrow})
   \big[\left(S_{2n-1}^{x}S_{2n}^{y} - S_{2n-1}^{y}S_{2n}^{x}\right)-
   \left(S_{2n}^{x}S_{2n+1}^{y} - S_{2n}^{y}S_{2n+1}^{x}\right)\big] +  \nonumber\\
   && \hspace{28mm} +  2 t^2_{\uparrow}t^2_{\downarrow}\left({\bf S}_{2n-1}
   \cdot {\bf S}_{2n+1}-1/4\right)\,\Big]\, ,\label{4T-A5}\\
%%%%%%%%%%%%%%%%%%%%%%%%%%%%%
%%%%%%%%%%%%%%%%%%%%%%%%%%%%%%%
&& \quad  T^{e}_{-1}T^{pe}_{0}T^{de}_{0}T^{o}_{1} = \sum_{n}\Big[\,-
t_{\uparrow}t_{\downarrow}(t^2_{\uparrow}+t^2_{\downarrow})
\big(S_{n}^{x}S_{n+1}^{x}+ S_{n}^{y}S_{n+1}^{y}\big) - 2
t^2_{\uparrow}t^2_{\downarrow} \left( S^{z}_{n}S^{z}_{n+1} -
1/4 \right) -  \nonumber\\
&&\hspace{28mm} - i\,
t_{\uparrow}t_{\downarrow}(t^2_{\uparrow}-t^2_{\downarrow})
\big[\left(S_{2n}^{x}S_{2n+1}^{y} - S_{2n}^{y}S_{2n+1}^{x}\right)
-\left(S_{2n+1}^{x}S_{2n+2}^{y} - S_{2n+1}^{y}S_{2n+2}^{x}\right)\big] +  \nonumber\\
   && \hspace{28mm} + 2 t^2_{\uparrow}t^2_{\downarrow}
   \left({\bf S}_{2n} \cdot {\bf S}_{2n+2}-1/4\right)\,\Big]\, ,\label{4T-A6}\\
%%%%%%%%%%%%%%%%%%%%%%%%%%%%%%%
%%%%%%%%%%%%%%%%%%%%%%%%%%%%%%%
&& \quad  T^{o}_{-1}T^{do}_{0}T^{po}_{0}T^{e}_{1} = \sum_{n}\Big[\,-
   t_{\uparrow}t_{\downarrow}(t^2_{\uparrow}+t^2_{\downarrow})
   \big(S_{n}^{x}S_{n+1}^{x}+ S_{n}^{y}S_{n+1}^{y}\big) -
   2 t^2_{\uparrow}t^2_{\downarrow}\left( S^{z}_{n}S^{z}_{n+1} - 1/4 \right) +  \nonumber\\  \nonumber\\
   && \hspace{28mm}+  i\, t_{\uparrow}t_{\downarrow}(t^2_{\uparrow}-t^2_{\downarrow})
   \big[\left(S_{2n}^{x}S_{2n+1}^{y} - S_{2n}^{y}S_{2n+1}^{x}\right)-
   \left(S_{2n+1}^{x}S_{2n+2}^{y} - S_{2n+1}^{y}S_{2n+2}^{x}\right)\big] +  \nonumber\\
   && \hspace{28mm}+  2 t^2_{\uparrow}t^2_{\downarrow}
   \left({\bf S}_{2n} \cdot {\bf S}_{2n+2}-1/4\right)\,\Big]\, ,\label{4T-A7}\\
%%%%%%%%%%%%%%%%%%%%%%%%%%%%%%%%%%%%%
%%%%%%%%%%%%%%%%%%%%%%%%%%%%%%%%%%%%%
&& \quad  T^{o}_{-1}T^{po}_{0}T^{do}_{0}T^{e}_{1} = \sum_{n}\Big[\,-
   t_{\uparrow}t_{\downarrow}(t^2_{\uparrow} +
   t^2_{\downarrow})\big(S_{n}^{x}S_{n+1}^{x}+ S_{n}^{y}S_{n+1}^{y}\big)
   - 2 t^2_{\uparrow}t^2_{\downarrow}\left( S^{z}_{n}S^{z}_{n+1} - 1/4 \right)
   -  \nonumber\\
   && \hspace{28mm} - i\, t_{\uparrow}t_{\downarrow}(t^2_{\uparrow}-t^2_{\downarrow})
   \big[\left(S_{2n-1}^{x}S_{2n}^{y} - S_{2n-1}^{y}S_{2n}^{x}\right)-
   \left(S_{2n}^{x}S_{2n+1}^{y} - S_{2n}^{y}S_{2n+1}^{x}\right)\big] +  \nonumber\\
   && \hspace{28mm} + 2 t^2_{\uparrow}t^2_{\downarrow}
   \left({\bf S}_{2n-1} \cdot {\bf
   S}_{2n+1}-1/4\right)\,\Big]\, .\label{4T-A8}
\end{eqnarray}
%%%%%%%%%%%%%%%%%%%%%%%%%%%%%

\subsubsection{Group B1: Four-{\it T} product terms of \texorpdfstring{the form
$T^{a}_{-1}T^{b}_{-1}T^{b}_{1}T^{a}_{1}$ and
$T^{b}_{-1}T^{a}_{-1}T^{b}_{1}T^{a}_{1}$\;($a \neq b$)}{type 2}}

There are four terms of this type in the  effective Hamiltonian
(\ref{Effect-Ham-In-T-Oper-4}). These terms correspond to processes
where two doublons are created on even and odd sites and then
consecutively annihilated. If the first pair is created on site
$\ell$, the hopping processes leading to the creation of the second
pair are restricted by the existence of an empty site adjacent
to $\ell$. Consequently, using the relations
(\ref{T00withX})-\!-(\ref{SpinXoperators}) one obtains a
restricted double summation of the form
%%%%%%%%%%%%%%%%%%%%%%%%%%%%%%%
\begin{eqnarray}
&T^{o}_{-1}T^{e}_{-1}T^{o}_{1}T^{e}_{1} =
\sum\limits_{n,m\not=n,n\pm1} %\sum_{m\not=n,n\pm1}
\Big[   - 2 t_{\uparrow}
t_{\downarrow}( S^{x}_{m} S^{x}_{m+1} +  S^{y}_{m} S^{y}_{m+1})-
    (t^{2}_{\uparrow}+t^{2}_{\downarrow})( S^{z}_{m}
    S^{z}_{m+1}-1/4)+
    \frac{t^{2}_{\uparrow}-t^{2}_{\downarrow}}{2}(-1)^{m}(S^{z}_{m}-S^{z}_{m+1})\Big]
     \cdot &\nonumber\\
&\hspace{25mm}\cdot\Big[- 2 t_{\uparrow} t_{\downarrow}( S^{x}_{n}
S^{x}_{n+1} +  S^{y}_{n} S^{y}_{n+1})-
    (t^{2}_{\uparrow}+t^{2}_{\downarrow})( S^{z}_{n}
    S^{z}_{n+1}-1/4)-
    \frac{t^{2}_{\uparrow}-t^{2}_{\downarrow}}{2}(-1)^{n}(S^{z}_{n}-S^{z}_{n+1})\Big]
    &\nonumber\\
    &\hspace{25mm}\equiv (T^{o}_{-1}T^{o}_{1})\ast (T^{e}_{-1}T^{e}_{1})\, .&
    \label{4T-B1-1}
\end{eqnarray}
%%%%%%%%%%%%%%%%%%%%%%%%%%%%%%%
Here we have introduced the notation $\ast$ to denote multiplication
of infinite sums over the indices $n$ and $m$ with the restrictive
condition $m\not=n,n\pm1$.

For the other terms of the same group we analogously obtain
%%%%%%%%%%%%%%%%%%%%%%%%%%%%%%%
\begin{eqnarray}
&& T^{e}_{-1}T^{o}_{-1}T^{o}_{1}T^{e}_{1}=(T^{o}_{-1}T^{o}_{1})\ast
(T^{e}_{-1}T^{e}_{1}) \, ,\label{4T-B1-2}\\
&&T^{e}_{-1}T^{o}_{-1}T^{e}_{1}T^{o}_{1}=T^{o}_{-1}T^{e}_{-1}T^{e}_{1}T^{o}_{1}
=(T^{e}_{-1}T^{e}_{1})\ast (T^{o}_{-1}T^{o}_{1})\, .\label{4T-B1-3}
\end{eqnarray}
%%%%%%%%%%%%%%%%%%%%%%%%%%%%%
However, since all the hopping processes in these products take
place on disjoint pairs of sites $(n,n+1)$ and $(m,m+1)$, one can
freely commute the $S$-operators past each other, so that the
order of the multiplicands becomes irrelevant:
%%%%%%%%%%%%%%%%%%%%%%%%%%%%%%%
\begin{equation}
(T^{o}_{-1}T^{o}_{1})\ast(T^{e}_{-1}T^{e}_{1})=(T^{e}_{-1}T^{e}_{1})\ast (T^{o}_{-1}T^{o}_{1})
=\frac{1}{2}\big[(T^{o}_{-1}T^{o}_{1})\ast(T^{e}_{-1}T^{e}_{1})+(T^{e}_{-1}T^{e}_{1})\ast (T^{o}_{-1}T^{o}_{1})\big]\, .\label{4T-B1-4}
\end{equation}
%%%%%%%%%%%%%%%%%%%%%%%%%%%%%%%

\subsubsection{Group B2: Four-{\it T} product terms of \texorpdfstring{the form
$T^{a}_{-1}T^{a}_{-1}T^{a}_{1}T^{a}_{1}$}{type 3}}

There are two terms of this type in the  effective Hamiltonian
(\ref{Effect-Ham-In-T-Oper-4}). These terms correspond to processes
where two doubly occupied sites are created either on even or on odd
sites and then consecutively annihilated.  As before, the creation
of the first pair puts limitations on the processes responsible for
the creation of the second pair. In addition, since there are two
different ways how one can get the same configuration corresponding
to the pair of doublons located on two odd or two even sites, an
extra factor of 2 appears in the expressions for these terms:
%%%%%%%%%%%%%%%%%%%%%%%%%%%%%%%%%%%%%
\begin{eqnarray}
T ^{o}_{-1}T^{o}_{-1}T^{o}_{1}T^{o}_{1} = \hspace{-15mm} && \nonumber \\
&=2\sum\limits_{n,m\not=n,n\pm1}\Big[   - 2 t_{\uparrow}
t_{\downarrow}( S^{x}_{m} S^{x}_{m+1} +  S^{y}_{m} S^{y}_{m+1})-
    (t^{2}_{\uparrow}+t^{2}_{\downarrow})( S^{z}_{m} S^{z}_{m+1}-1/4)+
    \frac{t^{2}_{\uparrow}-t^{2}_{\downarrow}}{2}(-1)^{m}(S^{z}_{m}-S^{z}_{m+1})\Big] \cdot& \nonumber\\
& \hspace{15mm}\cdot\Big[- 2 t_{\uparrow} t_{\downarrow}( S^{x}_{n}
S^{x}_{n+1} +  S^{y}_{n} S^{y}_{n+1})-
    (t^{2}_{\uparrow}+t^{2}_{\downarrow})( S^{z}_{n} S^{z}_{n+1}-1/4)+
    \frac{t^{2}_{\uparrow}-t^{2}_{\downarrow}}{2}(-1)^{n}(S^{z}_{n}-S^{z}_{n+1})\Big]&\nonumber\\
    &\equiv 2 (T^{o}_{-1}T^{o}_{1})\ast (T^{o}_{-1}T^{o}_{1})\, ,&   \label{4T-B2-1}
\end{eqnarray}
%%%%%%%%%%%%%%%%%%%%%%%%%%%%%%%%%%%%%
\begin{eqnarray}
T ^{e}_{-1}T^{e}_{-1}T^{e}_{1}T^{e}_{1} = \hspace{-15mm} && \nonumber \\
&=2\sum\limits_{n,m\not=n,n\pm1}\Big[   - 2 t_{\uparrow}
t_{\downarrow}( S^{x}_{m} S^{x}_{m+1} +  S^{y}_{m} S^{y}_{m+1})-
    (t^{2}_{\uparrow}+t^{2}_{\downarrow})( S^{z}_{m} S^{z}_{m+1}-1/4)-
    \frac{t^{2}_{\uparrow}-t^{2}_{\downarrow}}{2}(-1)^{m}(S^{z}_{m}-S^{z}_{m+1})\Big]
    \cdot &\nonumber\\
& \hspace{15mm}\cdot\Big[- 2 t_{\uparrow} t_{\downarrow}( S^{x}_{n}
S^{x}_{n+1} +  S^{y}_{n} S^{y}_{n+1})-
    (t^{2}_{\uparrow}+t^{2}_{\downarrow})( S^{z}_{n}
    S^{z}_{n+1}-1/4)-
    \frac{t^{2}_{\uparrow}-t^{2}_{\downarrow}}{2}(-1)^{n}(S^{z}_{n}-S^{z}_{n+1})\Big]&\nonumber\\
    &\equiv 2(T^{e}_{-1}T^{e}_{1})\ast (T^{e}_{-1}T^{e}_{1})\, .&   \label{4T-B2-2}
\end{eqnarray}
%%%%%%%%%%%%%%%%%%%%%%%%%%%%%%%%%%%%%%%%%%%%%%%%%%%%%%%%%

\subsubsection{Group C: Four-{\it T} product terms of \texorpdfstring{the form
$T_{-1}T_{1}T_{-1}T_{1}$}{type 4}}

There are four terms of this type in the  effective Hamiltonian
(\ref{Effect-Ham-In-T-Oper-4}). These terms correspond to processes
where a doublon is created and immediately annihilated on a site
$\ell$ and then another doublon is created and annihilated on an
arbitrary  site $\ell^{\prime}$. Using
(\ref{T00withX})-\!-(\ref{SpinXoperators}), we obtain
%%%%%%%%%%%%%%%%%%%%%%%%%%%%%%%%%%%%%
\begin{eqnarray}
&T ^{o}_{-1}T^{o}_{1}T^{o}_{-1}T^{o}_{1} =\sum\limits_{n,m}\Big[ - 2
t_{\uparrow} t_{\downarrow}( S^{x}_{m} S^{x}_{m+1} +  S^{y}_{m}
S^{y}_{m+1})-
    (t^{2}_{\uparrow}+t^{2}_{\downarrow})( S^{z}_{m}
    S^{z}_{m+1}-1/4)+
    \frac{t^{2}_{\uparrow}-t^{2}_{\downarrow}}{2}(-1)^{m}(S^{z}_{m}-S^{z}_{m+1})\Big]
    \cdot& \nonumber\\
& \hspace{35mm}\cdot\Big[- 2 t_{\uparrow} t_{\downarrow}( S^{x}_{n}
S^{x}_{n+1} +  S^{y}_{n} S^{y}_{n+1})-
    (t^{2}_{\uparrow}+t^{2}_{\downarrow})( S^{z}_{n} S^{z}_{n+1}-1/4)+
    \frac{t^{2}_{\uparrow}-t^{2}_{\downarrow}}{2}(-1)^{n}(S^{z}_{n}-S^{z}_{n+1})\Big]&\nonumber\\
    &\hspace{25mm}\equiv  (T^{o}_{-1}T^{o}_{1})\cdot (T^{o}_{-1}T^{o}_{1})\,
    ,&
    \label{4T-C-1}
\end{eqnarray}
%%%%%%%%%%%%%%%%%%%%%%%%%%%%%%%%%%%%%
%%%%%%%%%%%%%%%%%%%%%%%%%%%%%%%%%%%%%%%%%%%%%%%%%%%%%%%%%%%%%%%%%%
\begin{eqnarray}
&T ^{e}_{-1}T^{e}_{1}T^{e}_{-1}T^{e}_{1} =\sum\limits_{n,m}\Big[ - 2
t_{\uparrow} t_{\downarrow}( S^{x}_{m} S^{x}_{m+1} +  S^{y}_{m}
S^{y}_{m+1})-
    (t^{2}_{\uparrow}+t^{2}_{\downarrow})( S^{z}_{m}
    S^{z}_{m+1}-1/4)-
    \frac{t^{2}_{\uparrow}-t^{2}_{\downarrow}}{2}(-1)^{m}(S^{z}_{m}-S^{z}_{m+1})\Big] \cdot& \nonumber\\
& \hspace{35mm}\cdot\Big[- 2 t_{\uparrow} t_{\downarrow}( S^{x}_{n}
S^{x}_{n+1} +  S^{y}_{n} S^{y}_{n+1})-
    (t^{2}_{\uparrow}+t^{2}_{\downarrow})( S^{z}_{n}
    S^{z}_{n+1}-1/4)-
    \frac{t^{2}_{\uparrow}-t^{2}_{\downarrow}}{2}(-1)^{n}(S^{z}_{n}-S^{z}_{n+1})\Big]&\nonumber\\
    &\hspace{25mm}\equiv
    (T^{e}_{-1}T^{e}_{1})\cdot(T^{e}_{-1}T^{e}_{1})\,&
    \label{4T-C-2}
\end{eqnarray}
%%%%%%%%%%%%%%%%%%%%%%%%%%%%%%%%%%%%%
and
%%%%%%%%%%%%%%%%%%%%%%%%%%%%%%%%%%%%%%%%%%%%%%%%%%%%%%%%%%%%%%%%%%
%%%%%%%%%%%%%%%%%%%%%%%%%%%%%%%%%%%%%
\begin{equation}
T ^{o}_{-1}T^{o}_{1}T^{e}_{-1}T^{e}_{1} = (T^{o}_{-1}T^{o}_{1})\cdot
(T^{e}_{-1}T^{e}_{1})\, , \qquad T^{e}_{-1}T^{e}_{1}T^{o}_{-1}T^{o}_{1} = (T^{e}_{-1}T^{e}_{1})\cdot
(T^{o}_{-1}T^{o}_{1})\, .  \label{4T-C-3}
\end{equation}
%%%%%%%%%%%%%%%%%%%%%%%%%%%%%%%%%%%%%
%%%%%%%%%%%%%%%%%%%%%%%%%%%%%%%%%%%%%%%%%%%%%%%%%%%%%%%%%%%%%%%%%%

\vspace{5mm}

As we observe, in marked contrast with the first eight terms
(\ref{4T-A1})-\!-(\ref{4T-A8}) which only couple spins located on
neighboring sites, the remaining ten terms given by
(\ref{4T-B1-1})-\!-(\ref{4T-C-3}) contain countless number of all
possible two-spin $S^{z}_{n}S^{z}_{m}$, three-spin $S^{p}_{n}
S^{p}_{n+1}S^{z}_{m}$ and four-spin $S^{p}_{n} S^{p}_{n+1}S^{q}_{m}
S^{q}_{m+1}$ combinations, where $p,q=x,y,z$. The situation is
rescued by the fact that after combining identical terms in the
Hamiltonian (\ref{Effect-Ham-In-T-Oper-4}), each term of the type
$(T^{a}_{-1}T^{a}_{1})\ast (T^{b}_{-1}T^{b}_{1})$ will have its
counterpart of the type $(T^{a}_{-1}T^{a}_{1})\cdot (T^{b}_{-1}T^{b}_{1})$
with just the opposite coefficient. As a result, all terms
corresponding to distant spin-spin interaction are canceled:
%%%%%%%%%%%%%%%%%%%%%%%%%%%%%%%%%%%%%%%%%%%%%%%%%%%%%%%%%%%%%%%%%%
\begin{eqnarray}
&(T^{o}_{-1}T^{o}_{1})\cdot
(T^{o}_{-1}T^{o}_{1})-(T^{o}_{-1}T^{o}_{1})\ast
(T^{o}_{-1}T^{o}_{1})=&\nonumber\\
&\hspace{10mm}=\sum\limits_{n,m=n,n\pm1}\Big[   - 2 t_{\uparrow}
t_{\downarrow}( S^{x}_{m} S^{x}_{m+1} +  S^{y}_{m} S^{y}_{m+1})-
    (t^{2}_{\uparrow}+t^{2}_{\downarrow})( S^{z}_{m}
    S^{z}_{m+1}-1/4)+
    \frac{t^{2}_{\uparrow}-t^{2}_{\downarrow}}{2}(-1)^{m}(S^{z}_{m}-S^{z}_{m+1})\Big] \cdot& \nonumber\\
& \hspace{30mm}\cdot\Big[- 2 t_{\uparrow} t_{\downarrow}( S^{x}_{n}
S^{x}_{n+1} +  S^{y}_{n} S^{y}_{n+1})-
    (t^{2}_{\uparrow}+t^{2}_{\downarrow})( S^{z}_{n} S^{z}_{n+1}-1/4)+
    \frac{t^{2}_{\uparrow}-t^{2}_{\downarrow}}{2}(-1)^{n}(S^{z}_{n}-S^{z}_{n+1})\Big]=&\nonumber\\
\nonumber\\
%%%%%%%%%%%%%%%%%%%%                                                          1-st part
 &\hspace{10mm}=\sum\limits_{n}\Big\{ \Big[-4 t_{\uparrow}
t_{\downarrow}(t^2_{\uparrow}+ t^2_{\downarrow})
    ( S^{x}_{n} S^{x}_{n+1} +  S^{y}_{n} S^{y}_{n+1})-
    (3 t^4_{\uparrow} + 2 t^2_{\uparrow}t^2_{\downarrow} +3 t^4_{\downarrow})
    ( S^{z}_{n} S^{z}_{n+1}-1/4) +& \nonumber\\
&\hspace{15mm}+  2 t^2_{\uparrow}t^2_{\downarrow}
    ( S^{x}_{n} S^{x}_{n+2} +  S^{y}_{n} S^{y}_{n+2})+
    (t^4_{\uparrow} + t^4_{\downarrow})( S^{z}_{n} S^{z}_{n+2}-1/4)\Big] +
    (-1)^{n}\Big[\,\frac{t^{4}_{\uparrow}-t^{4}_{\downarrow}}{2}
    (4 S^{z}_{n-1}S^{z}_{n}S^{z}_{n+1} + 5 S^{z}_{n})&\nonumber\\
&\hspace{15mm}  + 2 t_{\uparrow}
t_{\downarrow}(t^{2}_{\uparrow}-t^{2}_{\downarrow})
    [S^{z}_{n-1}( S^{x}_{n} S^{x}_{n+1} +  S^{y}_{n} S^{y}_{n+1}) +
    ( S^{x}_{n-1} S^{x}_{n} +  S^{y}_{n-1} S^{y}_{n})S^{z}_{n+1}] \, \Big]\Big\}\,
    ,&
    \label{TdotT-TastT-1}
\end{eqnarray}
%%%%%%%%%%%%%%%%%%%%%%%%%%%%%%%%%%%%%%%%%%%%%%%%%%%%%%%%%
%%%%%%%%%%%%%%%%%%%%%%%%%%%%%%%%%%%%%%%%%%%%%%%%%%%%%%%%%%%%%%%%%%
\begin{eqnarray}
&(T^{e}_{-1}T^{e}_{1})\cdot
(T^{e}_{-1}T^{e}_{1})-(T^{e}_{-1}T^{e}_{1})\ast
(T^{e}_{-1}T^{e}_{1})=&\nonumber\\
&\hspace{10mm}=\sum\limits_{n,m=n,n\pm1}\Big[   - 2 t_{\uparrow}
t_{\downarrow}( S^{x}_{m} S^{x}_{m+1} +  S^{y}_{m} S^{y}_{m+1})-
    (t^{2}_{\uparrow}+t^{2}_{\downarrow})( S^{z}_{m} S^{z}_{m+1}-1/4)-
    \frac{t^{2}_{\uparrow}-t^{2}_{\downarrow}}{2}(-1)^{m}(S^{z}_{m}-S^{z}_{m+1})\Big] \cdot& \nonumber\\
& \hspace{25mm}\cdot\Big[- 2 t_{\uparrow} t_{\downarrow}( S^{x}_{n}
S^{x}_{n+1} +  S^{y}_{n} S^{y}_{n+1})-
    (t^{2}_{\uparrow}+t^{2}_{\downarrow})( S^{z}_{n}
    S^{z}_{n+1}-1/4)-
    \frac{t^{2}_{\uparrow}-t^{2}_{\downarrow}}{2}(-1)^{n}(S^{z}_{n}-S^{z}_{n+1})\Big]=&\nonumber\\
    %%%                                                                     1-st line
&\hspace{10mm}=\sum\limits_{n}\Big\{ \Big[-4 t_{\uparrow}
t_{\downarrow}(t^2_{\uparrow}+ t^2_{\downarrow})
    ( S^{x}_{n} S^{x}_{n+1} +  S^{y}_{n} S^{y}_{n+1})-
    (3 t^4_{\uparrow} + 2 t^2_{\uparrow}t^2_{\downarrow} +3 t^4_{\downarrow})
    ( S^{z}_{n} S^{z}_{n+1}-1/4) + &\nonumber\\
&\hspace{15mm}+  2 t^2_{\uparrow}t^2_{\downarrow}
    ( S^{x}_{n} S^{x}_{n+2} +  S^{y}_{n} S^{y}_{n+2})+
    (t^4_{\uparrow} + t^4_{\downarrow})( S^{z}_{n} S^{z}_{n+2}-1/4)\Big]
    -
    (-1)^{n}\Big[\,\frac{t^{4}_{\uparrow}-t^{4}_{\downarrow}}{2}
    (4 S^{z}_{n-1}S^{z}_{n}S^{z}_{n+1} + 5 S^{z}_{n})&\nonumber\\
&\hspace{15mm}  + 2 t_{\uparrow}
t_{\downarrow}(t^{2}_{\uparrow}-t^{2}_{\downarrow})
    [S^{z}_{n-1}( S^{x}_{n} S^{x}_{n+1} +  S^{y}_{n} S^{y}_{n+1}) +
    ( S^{x}_{n-1} S^{x}_{n} +  S^{y}_{n-1} S^{y}_{n})S^{z}_{n+1}] \, \Big]\Big\}\,&
    \label{TdotT-TastT-2}
\end{eqnarray}
%%%%%%%%%%%%%%%%%%%%%%%%%%%%%%%%%%%%%%%%%%%%%%%%%%%%%%%%%
and
%%%%%%%%%%%%%%%%%%%%%%%%%%%%%%%%%%%%%%%%%%%%%%%%%%%%%%%%%%%%%%%%%
\begin{eqnarray}
&&\frac{1}{2}\left[(T^{o}_{-1}T^{o}_{1})\cdot
(T^{e}_{-1}T^{e}_{1})-(T^{o}_{-1}T^{o}_{1})\ast
(T^{e}_{-1}T^{e}_{1})\right] + \frac{1}{2}\left[(T^{e}_{-1}T^{e}_{1})\cdot
(T^{o}_{-1}T^{o}_{1})-(T^{e}_{-1}T^{e}_{1})\ast
(T^{o}_{-1}T^{o}_{1})\right]=\nonumber\\
&&\hspace{25mm}=  \sum_{n} \Big\{-4 t_{\uparrow}
t_{\downarrow}(t^2_{\uparrow}+ t^2_{\downarrow})
    ( S^{x}_{n} S^{x}_{n+1} +  S^{y}_{n} S^{y}_{n+1})-
    8 t^2_{\uparrow}t^2_{\downarrow} ( S^{z}_{n} S^{z}_{n+1}-\frac{1}{4}) + \nonumber\\
&&\hspace{40mm}+  2 t^2_{\uparrow}t^2_{\downarrow}
    ( S^{x}_{n} S^{x}_{n+2} +  S^{y}_{n} S^{y}_{n+2})+
    2 t^2_{\uparrow}t^2_{\downarrow}( S^{z}_{n} S^{z}_{n+2}-\frac{1}{4})\Big\}\,
    .
    \label{TdotT-TastT-3}
\end{eqnarray}
%%%%%%%%%%%%%%%%%%%%%%%%%%%%%%%%%%%%%%%%%%%%%%%%%%%%%%%%%%%%%%%%%%
For the last two terms we have taken their combination to avoid
calculation of extra terms which cancel each other in the sum.

Inserting finally the relations (\ref{4T-A1})-\!-(\ref{TdotT-TastT-3})
into (\ref{Effect-Ham-In-T-Oper-4}) we obtain that up to the fourth
order the strong-coupling effective spin Hamiltonian for the
spin-asymmetric alternating-U ionic Hubbard model is given by
%%%%%%%%%%%%%%%%%%%%%%%%%%%%%%%%%%%%%%%%%%%%%%%%%%%%%%%%%
\begin{eqnarray}\label{EffectiveHeisChainHam-2}
{\cal H}^{(4)}_{eff} & =& -\frac{T^{o}_{-1}T^{o}_{1}}{U_{o} - \Delta}
-\frac{T^{e}_{-1}T^{e}_{1}}{U_{e} + \Delta}-   \nonumber \\
%%%%%%%%%%%%%%%%%%                                                line 1
&& - \; \frac{1}{U_{o}}\left[\frac{T^{o}_{-1}T^{po}_{0}T^{pe}_{0}T^{o}_{1}}{(U_{o} - \Delta
)^{2}}+\frac{T^{e}_{-1}T^{de}_{0}T^{do}_{0}T^{e}_{1}}{(U_{e}
+ \Delta)^{2}}\right] -\frac{1}{U_{e}}\left[\frac{T^{o}_{-1}T^{do}_{0}T^{de}_{0}T^{o}_{1}}{(U_{o}
-\Delta)^{2}}+\frac{T^{e}_{-1}T^{pe}_{0}T^{po}_{0}T^{e}_{1}}{(U_{e} +
\Delta )^{2}}\right]-\nonumber \\
%%%%%%%%%%%%%%%%%%                                                line 2
&& - \; \frac{1}{(U_{o} - \Delta )U_{o}(U_{e} + \Delta)}
\left[T^{o}_{-1}T^{po}_{0}T^{do}_{0}T^{e}_{1}+
T^{e}_{-1}T^{de}_{0}T^{pe}_{0}T^{o}_{1}\right]- \nonumber \\
%%%%%%%%%%%%%%%%%%                                                line 3
&& - \; \frac{1}{(U_{o} - \Delta)U_{e}(U_{e} + \Delta)}
\left[T^{o}_{-1}T^{do}_{0}T^{po}_{0}T^{e}_{1}+
T^{e}_{-1}T^{pe}_{0}T^{de}_{0}T^{o}_{1}\right]+ \nonumber \\
%%%%%%%%%                                                          line 4
&& + \; \frac{1}{(U_{o}
- \Delta )^{3}}\left[(T^{o}_{-1}T^{o}_{1})\cdot
(T^{o}_{-1}T^{o}_{1})-(T^{o}_{-1}T^{o}_{1})\ast
(T^{o}_{-1}T^{o}_{1})\right]+ \nonumber \\
%%%%%%%%%                                                          line 5
&& + \; \frac{1}{(U_{e}
+ \Delta )^{3}}\left[(T^{e}_{-1}T^{e}_{1})\cdot
(T^{e}_{-1}T^{e}_{1})-(T^{e}_{-1}T^{e}_{1})\ast
(T^{e}_{-1}T^{e}_{1})\right]+ \nonumber \\
%%%%%%%%%                                                          line 6
&& + \; \frac{U_{o}+U_{e}}{(U_{o} - \Delta )^{2}(U_{e} + \Delta)^{2}}\Big\{\, \frac{1}{2}\left[(T^{o}_{-1}T^{o}_{1})\cdot
(T^{e}_{-1}T^{e}_{1})-(T^{o}_{-1}T^{o}_{1})\ast
(T^{e}_{-1}T^{e}_{1})\right] + \nonumber \\
%%%%%%%%%                                                          line 7
&& + \; \frac{1}{2}\left[(T^{e}_{-1}T^{e}_{1})\cdot
(T^{o}_{-1}T^{o}_{1})-(T^{e}_{-1}T^{e}_{1})\ast
(T^{o}_{-1}T^{o}_{1})\right]\Big\}= \nonumber \\
%%%%%%%%%%%%%%%%%%                                                 line 8
& = & \sum_{n} \Big[ J_{\perp} ( S^{x}_{n}
S^{x}_{n+1} + S^{y}_{n} S^{y}_{n+1}) + J_{\parallel}( S^{z}_{n}
S^{z}_{n+1}-\frac{1}{4}) \Big] - \sum_{n} h(n)\,S^{z}_{n}
\nonumber\\
%%%%%%%                                                             Line 1
&& + \; \sum_{n}\Big[J^{\prime}_{\perp}(n)\,( S^{x}_{n} S^{x}_{n+2} +
S^{y}_{n} S^{y}_{n+2})+ J^{\prime}_{\parallel}(n)\,( S^{z}_{n}
S^{z}_{n+2}-\frac{1}{4})\Big]
\nonumber\\
%%%%%%%                                                             Line 2
&& + \; \sum_{n}\Big[ W_{\perp}(n)\,[( S^{x}_{n-1} S^{x}_{n} +
S^{y}_{n-1} S^{y}_{n}) S^{z}_{n+1}+
 S^{z}_{n-1}( S^{x}_{n} S^{x}_{n+1} +  S^{y}_{n} S^{y}_{n+1})] +
 W_{\parallel}(n)\, S^{z}_{n-1} S^{z}_{n} S^{z}_{n+1}\Big]  \; ,
%%%%%%%                                                             Line 3
\end{eqnarray}
%%%%%%%%%%%%%%%%%%%%%%%%%%%%%%%%%%%%%%%%%%%%%%%%%%%%%%%%%%%%%%%%%%
where
%%%%%%%%%%%%%%%%%%%%%%%%%%%%%%%%%%%%%%%%%%%%%%%%%%%%%%%%%%%%%%%%%
\begin{eqnarray}
J_{\perp} & = & 2 t_{\uparrow} t_{\downarrow} \left[\frac{1}{U_{o} -
\Delta } +
\frac{1}{U_{e} + \Delta}\right]+  \nonumber\\
%%%%%%%%                                                                     Line 1
   && + \; t_{\uparrow} t_{\downarrow}(t^{2}_{\uparrow}+t^{2}_{\downarrow})
   \frac{U_{o}+U_{e}}{U_{o}U_{e}}
   \Big[\frac{1}{(U_{o} -  \Delta)^2 }+\frac{1}{(U_{e} +  \Delta)^2 }
    + \frac{2}{(U_{o} -  \Delta)(U_{e} + \Delta )}\Big]-   \nonumber\\
%%%%%%%%                                                                     Line 2
   && -  \; 4t_{\uparrow} t_{\downarrow}(t^{2}_{\uparrow}+t^{2}_{\downarrow})
   \Big[\frac{1}{(U_{o} -  \Delta)^3}+ \frac{1}{(U_{e} +  \Delta)^{3}} +\frac{U_{o} + U_{e}}{(U_{o} -  \Delta)^2(U_{e} + \Delta )^2} \Big]=    \nonumber\\
%%%%%%%%                                                                     Line 3
\nonumber\\
& = &\frac{4t_{\uparrow} t_{\downarrow}}{U}\frac{1}{1-\lambda^{2}}+ \frac{t_{\uparrow}
t_{\downarrow}(t^{2}_{\uparrow}+t^{2}_{\downarrow})}{U^{3}}\frac{2}{(1-\delta^{2})}
\frac{4}{(1-\lambda^{2})^{2}}-
\frac{4t_{\uparrow}
t_{\downarrow}(t^{2}_{\uparrow}+t^{2}_{\downarrow})}{U^{3}}\frac{4(1+\lambda^{2})}
{(1-\lambda^{2})^{3}}= \nonumber\\
& = &\frac{4t_{\uparrow} t_{\downarrow}}{U(1-\lambda^{2})} \left[ 1
-\frac{2(t^{2}_{\uparrow}+t^{2}_{\downarrow})}{U^{2}(1-\lambda^{2})^{2}}
\left(2+ 2\lambda^{2}-
\frac{1-\lambda^{2}}{1-\delta^{2}}\right) \right] \; ,  \\
%%%%%%%%%%%%%%%%%%%%%%%%%%%%%%%%%%%%%%%%%%%%%%%%%%%%%%%%%%%%%%%%%%
\nonumber  \\
%%%%%%%%%%%%%%%%%%%%%%%%%%%%%%%%%%%%%%%%%%%%%%%%%%%%%%%%%%%%%%%%%%
J_{\parallel} & = &
(t^{2}_{\uparrow}+t^{2}_{\downarrow})\left[\frac{1}{U_{o} - \Delta }
+ \frac{1}{U_{e} + \Delta}\right]
 - 3(t^{4}_{\uparrow}+t^{4}_{\downarrow})\left[
 \frac{1}{(U_{o} -  \Delta)^3} + \frac{1}{(U_{e} + \Delta)^{3}} \right]+ \nonumber\\
%%%%%%%%                                                                     Line 1
   && + \; 2 t^{2}_{\uparrow} t^{2}_{\downarrow} \frac{U_{o}+U_{e}}{U_{o}U_{e}}
   \Big[ \frac{1}{(U_{o} -  \Delta)^2} + \frac{1}{(U_{e} + \Delta)^2}
   + \frac{2}{(U_{o} -  \Delta)(U_{e} + \Delta )}\Big]-   \nonumber\\
%%%%%%%%                                                                     Line 2
   && - \; 2 t^{2}_{\uparrow} t^{2}_{\downarrow}\left[\frac{1}{(U_{o} -  \Delta)^3}
   + \frac{1}{(U_{e} +  \Delta)^{3}}
   + \frac{4 \, (U_{o}+U_{e})}{(U_{o} - \Delta)^{2}(U_{e} + \Delta )^{2}} \right]= \nonumber\\
   & = & \frac{2(t^{2}_{\uparrow}+t^{2}_{\downarrow})}{U(1-\lambda^{2})}-
   \frac{6(t^{4}_{\uparrow}+t^{4}_{\downarrow})}{U^{3}(1-\lambda^{2})^{3}}(1+3\lambda^{2}) -\frac{4t^{2}_{\uparrow}t^{2}_{\downarrow}}{U^{3}(1-\lambda^{2})^{3}}
   \left(5-\lambda^{2} - \frac{4(1-\lambda^{2})}{1-\delta^{2}}\right)  \; ,  \\
%%%%%%%%%%%%%%%%%%%%%%%%%%%%%%%%%%%%%%%%%%%%%%%%%%%%%%%%%%%%%%%%%%
\nonumber  \\
%%%%%%%%%%%%%%%%%%%%%%%%%%%%%%%%%%%%%%%%%%%%%%%%%%%%%%%%%%%%%%%%%%
h(n) & = & (-1)^{n} \Big \{(t^2_{\uparrow} - t^2_{\downarrow})\Big[\frac{1}{U_{o} - \Delta }
   - \frac{1}{U_{e} + \Delta}\Big] + \nonumber\\
   && + \; \frac{(t^{4}_{\uparrow}- t^{4}_{\downarrow})}{2}\frac{2}{( U_{e} + U_{o})-(-1)^{n} ( U_{e} - U_{o})} \Big[\frac{1}{(U_{o} - \Delta )^{2}} - \frac{1}{(U_{e} + \Delta)^2}\Big] - \nonumber \\
   && - \; \frac{5 (t^{4}_{\uparrow}- t^{4}_{\downarrow})}{2}\Big[\frac{1}{(U_{o} - \Delta )^{3}} - \frac{1}{(U_{e} + \Delta )^{3}}\Big] \Big \}=   \nonumber\\
   & = & (-1)^{n} \left\{ \frac{2\lambda(t^{2}_{\uparrow}-t^{2}_{\downarrow})}{U(1-\lambda^{2})}-\frac{\lambda(t^{4}_{\uparrow} - t^{4}_{\downarrow})}{U^{3}(1-\lambda^{2})^{3}} \left[5 \, (3+ \lambda^{2})-
   \frac{2(1-\lambda^{2})}{1-\delta^{2}}\right] \right \}+\frac{2\lambda\delta(t^{4}_{\uparrow} - t^{4}_{\downarrow})}{U^{3}(1-\lambda^{2})^{2}(1-\delta^{2})}  \; ,  \\
%%%%%%%%%%%%%%%%%%%%%%%%%%%%%%%%%%%%%%%%%%%%%%%%%%%%%%%%%%%%%%%%%%
\nonumber  \\
%%%%%%%%%%%%%%%%%%%%%%%%%%%%%%%%%%%%%%%%%%%%%%%%%%%%%%%%%%%%%%%%%%
J^{\prime}_{\perp}(n)\, & = & - \, 2 t^{2}_{\uparrow} t^{2}_{\downarrow} \,
\frac{4}{[(U_{e} + U_{o})+(-1)^{n}(U_{e} - U_{o})](U_{o} - \Delta)(U_{e} + \Delta )} + \nonumber\\
%%%%%%%                                                             Line 1
&& + \; 2 t^{2}_{\uparrow} t^{2}_{\downarrow}
\Big[ \frac{1}{(U_{o} -  \Delta)^3} + \frac{1}{(U_{e} + \Delta)^{3}} + \frac{U_{e} + U_{o}}{(U_{o} - \Delta)^{2}(U_{e} + \Delta )^{2}} \Big] =  \nonumber\\
& = & \frac{4t^{2}_{\uparrow} t^{2}_{\downarrow}}{U^{3}(1-\lambda^{2})^{3}} \left[2 + 2\lambda^{2}-
   \frac{(1-\lambda^{2})^2}{1-\delta^{2}}\right]+(-1)^{n}\frac{4\delta t^{2}_{\uparrow} t^{2}_{\downarrow}}{U^{3}(1-\lambda^{2})(1-\delta^{2})} \; ,  \\
%%%%%%%%%%%%%%%%%%%%%%%%%%%%%%%%%%%%%%%%%%%%%%%%%%%%%%%%%%%%%%%%%%
%%%%%%%%%%%%%%%%%%%%%%%%%%%%%%%%%%%%%%%%%%%%%%%%%%%%%%%%%%%%%%%%%%
J^{\prime}_{\parallel}(n)\, & = & (t^{2}_{\uparrow} -
t^{2}_{\downarrow})^{2}
\frac{2}{(U_{e} + U_{o})+(-1)^{n}(U_{e} - U_{o})} \Big[ \frac{1}{(U_{o} - \Delta)^{2}} + \frac{1}{(U_{e} + \Delta )^{2}} \Big] - \nonumber\\
%%%%%%%                                                             Line 1
&& - \; 2 t^{2}_{\uparrow} t^{2}_{\downarrow} \,
\frac{4}{[(U_{e} + U_{o})+(-1)^{n}(U_{e} - U_{o})](U_{o} - \Delta)(U_{e} + \Delta )} + \nonumber\\
%%%%%%%
&& + \; (t^{4}_{\uparrow} + t^{4}_{\downarrow})
\Big[ \frac{1}{(U_{o} -  \Delta)^3} + \frac{1}{(U_{e} + \Delta)^{3}} \Big] +  2 t^{2}_{\uparrow} t^{2}_{\downarrow}\frac{U_{e} + U_{o}}{(U_{o} - \Delta)^{2}(U_{e} + \Delta )^{2}}=  \nonumber\\
& = & - \, \frac{4t^{2}_{\uparrow} t^{2}_{\downarrow}(1+\delta^{2})}{U^{3}(1-\lambda^{2})^{2}(1-\delta^{2})} + \frac{2(t^{4}_{\uparrow} + t^{4}_{\downarrow})}{U^{3}(1-\lambda^{2})^{3}} \left[1+3\lambda^{2} + \frac{1 - \lambda^{4}}{1-\delta^{2}}\right]+ \nonumber \\
&& + \; (-1)^{n}\frac{2\delta}{U^{3}(1-\lambda^{2})^{2}(1-\delta^{2})}\big[4t^{2}_{\uparrow} t^{2}_{\downarrow}-(t^{4}_{\uparrow} + t^{4}_{\downarrow})(1+\lambda^{2})\big]  \; ,   \\
%%%%%%%%%%%%%%%%%%%%%%%%%%%%%%%%%%%%%%%%%%%%%%%%%%%%%%%%%%%%%%%%%%
\nonumber  \\
%%%%%%%%%%%%%%%%%%%%%%%%%%%%%%%%%%%%%%%%%%%%%%%%%%%%%%%%%%%%%%%%%%
W_{\perp}(n) & = & (-1)^{n} \Big \{
2t_{\uparrow}t_{\downarrow}(t^2_{\uparrow} - t^2_{\downarrow})
\frac{2}{( U_{e} + U_{o})-(-1)^{n} ( U_{e} - U_{o})} \Big[\frac{1}{(U_{o} - \Delta )^{2}} - \frac{1}{(U_{e} + \Delta)^2}\Big] + \nonumber \\
   && + \; 2t_{\uparrow}t_{\downarrow}(t^2_{\uparrow} - t^2_{\downarrow})\Big[\frac{1}{(U_{o} - \Delta )^{3}} - \frac{1}{(U_{e} + \Delta )^{3}}\Big] \Big \}= \nonumber\\
   & = & \frac{4\lambda \, t_{\uparrow}t_{\downarrow}(t^2_{\uparrow} - t^2_{\downarrow})}{U^{3}(1-\lambda^{2})^{2}} \left\{ \frac{2\delta}{1-\delta^{2}}+(-1)^{n} \left[ \frac{3+ \lambda^{2}}{1-\lambda^{2}}+ \frac{2}{1-\delta^{2}}\right] \right \}  \; ,  \\
%%%%%%%%%%%%%%%%%%%%%%%%%%%%%%%%%%%%%%%%%%%%%%%%%%%%%%%%%%%%%%%%%%
\nonumber  \\
%%%%%%%%%%%%%%%%%%%%%%%%%%%%%%%%%%%%%%%%%%%%%%%%%%%%%%%%%%%%%%%%%%
W_{\parallel}(n) & = & (-1)^{n} \Big \{ 2(t^4_{\uparrow} -
t^4_{\downarrow})
\frac{2}{( U_{e} + U_{o})-(-1)^{n} ( U_{e} - U_{o})} \Big[\frac{1}{(U_{o} - \Delta )^{2}} - \frac{1}{(U_{e} + \Delta)^2}\Big] + \nonumber \\
   && + \; 2(t^4_{\uparrow} - t^4_{\downarrow})\Big[\frac{1}{(U_{o} - \Delta )^{3}} - \frac{1}{(U_{e} + \Delta )^{3}}\Big] \Big \}= \nonumber\\
   & = & \frac{4\lambda (t^4_{\uparrow} - t^4_{\downarrow})}{U^{3}(1-\lambda^{2})^{2}} \left\{ \frac{2\delta}{1-\delta^{2}}+(-1)^{n} \left[ \frac{3+ \lambda^{2}}{1-\lambda^{2}}+ \frac{2}{1-\delta^{2}}\right] \right \} \; .  \label{Wparallel-2}  \\   \nonumber
\end{eqnarray}
%%%%%%%%%%%%%%%%%%%%%%%%%%%%%%%%%%%%%%%%%%%%%%%%%%%%%%%%%%%%%%%%%%

It should be pointed out that the presented expressions
(\ref{EffectiveHeisChainHam-2})-\!-(\ref{Wparallel-2}) fully
agree with the known results in the limiting cases
 of the standard \cite{MacDonald88} ($t_{\uparrow} =
t_{\downarrow}=t,\;\Delta=\delta=\lambda=0$) and the
alternating-$U$ \cite{AHM04} ($t_{\uparrow} =
t_{\downarrow}=t,\;\Delta=0,\;\lambda=\delta \neq 0$) Hubbard
models, but some of the fourth-order coefficients do not coincide
with the results obtained previously for the ionic \cite{Nagaosa86}
($t_{\uparrow} = t_{\downarrow}=t,\;\delta=0,\;\lambda=\Delta/U=x$)
and the spin-asymmetric \cite{Fath95} ($t_{\uparrow} \neq
t_{\downarrow},\;\Delta=\delta=\lambda=0$) Hubbard models. More
specifically, for the ionic Hubbard chain we arrive at a different
expression in the numerator of the nearest-neighbor coupling $J$,
whereas for the spin-asymmetric Hubbard model the disparities
concern the numerators of the coefficients $J_{\parallel}$ and
$J^{\prime}_{\parallel}$. We presume that these discrepancies are
due to the perturbative schemes adopted by the authors of
Refs.~\;~\cite{Nagaosa86} and~\;~\cite{Fath95}, as it is known that some
of these procedures are not sufficiently well-controlled at higher
orders. \cite{Oles90,MacDonald90}

\section{Conclusion} \label{concl}

We have derived the effective spin Hamiltonian for the low-energy
sector of the one-dimensional half-filled spin-asymmetric
alternating-$U$ ionic Hubbard model in the limit of strong on-site
repulsion. The obtained Hamiltonian is that of a frustrated
Heisenberg chain with alternating next-nearest-neighbor exchange and
three-spin coupling in the presence of a uniform and a staggered
magnetic field. As expected, the nnn exchange is larger for two
spins separated by a site with low on-site repulsion than for spins
separated by a site with high on-site repulsion. The intensity of
the three-spin coupling and the amplitudes of the magnetic fields
are proportional to the product of the parameter $\lambda$, which
reflects the broken translational symmetry of the lattice, and the
difference between up- and down-spin electron hopping amplitudes
$t_{\uparrow} - t_{\downarrow}$. The most dominant effect however
comes from the staggered magnetic field, and therefore, in marked
contrast with the spin-isotropic case $t_{\uparrow}=t_{\downarrow}$,
the ground-state properties of the considered electron system are
described by a spin-chain model with explicitly broken translational
symmetry.

We also remark that the general picture outlined above remains valid
in the case of a half-filled bipartite lattice of a higher dimension
-- to the lowest order one again obtains the anisotropic
nearest-neighbor spin exchange and the staggered magnetic field
which, as before, dominates any higher-order terms arising from the
more complex lattice geometry.

\section{Acknowledgments}

The authors would like to thank D.~Baeriswyl for careful reading of
the manuscript and many useful comments and suggestions. We
acknowledge support from the Swiss National Science Foundation
through the SCOPES grant No. IZ73Z0-128058. GIJ also acknowledges
support from the Georgian National Science Foundation and Science
and Technology Center in Ukraine through the joint grant No.
STCU-5906.

\appendix

\section*{Appendix: Derivation of the effective spin exchange expressions for the
 four-{\it T} product terms of \texorpdfstring{the form $T_{-1}T_{0}T_{0}T_{1}$}{type 1}}
\label{apndx}

In order to rewrite the products of four $T$-operators in the language of spin $S=1/2$ operators,
once again we make use of the Hubbard $X$-operators defined in section \ref{HubbardOperators};
the procedure is in essence the same as the one employed in section \ref{2ndOrder}
for the terms consisting of two $T$-operators.

Let us consider for example the term
$T^{o}_{-1}T^{po}_{0}T^{pe}_{0}T^{o}_{1}$, which belongs to the
group of the processes where the electron pair is created and
annihilated on the same site. In the summations below, the square
brackets around the lattice indices $[n, m, k]$ indicate that for an
even site $ 2m $ its odd partners $ 2k+1 $ and $ 2n+1 $ represent
neighboring sites. Thus, for a fixed $ m $ we have two possible
sets: $ k=m-1 $ and $ n=m $, or $ k=m $ and $ n=m-1 $.
%%%%%%%%%%%%%%%%%%%%%%%%%%%%%%%%%%%%%
\begin{eqnarray} \label{4T-A1-2}
&& \hspace{-12mm} T^{o}_{-1}T^{po}_{0}T^{pe}_{0}T^{o}_{1}= \nonumber \\
&=&\!\!\sum_{[n,m,k]}\sum_{\sigma}\eta(\sigma)\,t_{\sigma}\,
X_{2m}^{\sigma 0}X_{2k+1}^{\sigmabar 2} \,\cdot
\sum_{\beta}t^{2}_{\beta}\,X_{2n+1}^{\beta 0}X_{2m}^{0\beta}
X_{2m}^{\beta 0}X_{2n+1}^{0\beta} \,\cdot
\sum_{\alpha}\eta(\alpha)\, t_{\alpha}\, X_{2k+1}^{2 \alphabar}X_{2m}^{0\alpha} \nonumber \\
&=&\!\!\sum_{[n,m,k]}\,\sum_{\alpha,\beta,\sigma}
\eta(\alpha)\,\eta(\sigma)\,t_{\alpha}\,t_{\sigma}\,t^{2}_{\beta}\,
X_{2k+1}^{\sigmabar \alphabar}X_{2m}^{\sigma\alpha}X_{2n+1}^{\beta\beta}  \nonumber\\
&=&\!\!\sum_{[n,m,k]} \Big[
t^{2}_{\uparrow}\,X_{2k+1}^{\downarrow\downarrow}X_{2m}^{\uparrow\uparrow}+
t^{2}_{\downarrow}\,X_{2k+1}^{\uparrow\uparrow}X_{2m}^{\downarrow\downarrow}-
t_{\uparrow}t_{\downarrow} \Big(
X_{2k+1}^{\uparrow\downarrow}X_{2m}^{\downarrow\uparrow}+
X_{2k+1}^{\downarrow\uparrow}X_{2m}^{\uparrow\downarrow} \Big) \Big]
\cdot \Big[t^{2}_{\uparrow}\,X_{2n+1}^{\uparrow \uparrow} +
t^{2}_{\downarrow}\,X_{2n+1}^{\downarrow \downarrow} \Big] \nonumber\\
&=&\!\!\sum_{[n,m,k]} \bigg[
\big(t^{2}_{\uparrow}+t^{2}_{\downarrow}\big)
\Big(\frac{1}{4}-S^{z}_{2k+1}S^{z}_{2m}\Big) +
\frac{t^{2}_{\uparrow}-t^{2}_{\downarrow}}{2}
\big(S^{z}_{2m}-S^{z}_{2k+1}\big) - t_{\uparrow}t_{\downarrow}
\big(S^{+}_{2k+1}S^{-}_{2m}+S^{-}_{2k+1}S^{+}_{2m} \big) \bigg]
\cdot \nonumber \\ && \hspace{1cm} \cdot \bigg[
\frac{t^{2}_{\uparrow}+t^{2}_{\downarrow}}{2}+
\big(t^{2}_{\uparrow}-t^{2}_{\downarrow}\big)S^{z}_{2n+1} \bigg] \nonumber\\
&=&\!\!\sum_{[n,m,k]} \bigg[
\frac{\big(t^{2}_{\uparrow}+t^{2}_{\downarrow}\big)^2}{2} \Big(\frac{1}{4}-S^{z}_{2k+1}S^{z}_{2m}\Big) + \frac{\big(t^{2}_{\uparrow}-t^{2}_{\downarrow}\big)^2}{2} \Big(S^{z}_{2m}S^{z}_{2n+1}-\frac{1}{4}+\frac{1}{4}-S^{z}_{2k+1}S^{z}_{2n+1}\Big) +  \nonumber \\
&& \hspace{12mm} +\,\frac{t^{4}_{\uparrow}-t^{4}_{\downarrow}}{4} \big(S^{z}_{2m}-S^{z}_{2k+1}\big)+
   \big(t^{4}_{\uparrow}-t^{4}_{\downarrow}\big)
   \Big(\frac{1}{4}-S^{z}_{2k+1}S^{z}_{2m}\Big)S^{z}_{2n+1} - \nonumber \\
&& \hspace{12mm} -\,t_{\uparrow}t_{\downarrow}\big(t^{2}_{\uparrow}+t^{2}_{\downarrow}\big) \big(S^{x}_{2k+1}S^{x}_{2m}+S^{y}_{2k+1}S^{y}_{2m} \big) -
2t_{\uparrow}t_{\downarrow}\big(t^{2}_{\uparrow}-t^{2}_{\downarrow}\big) \big(S^{x}_{2k+1}S^{x}_{2m}+S^{y}_{2k+1}S^{y}_{2m} \big)S^{z}_{2n+1} \bigg]  \nonumber \\
&=& \sum_{\ell}  2 t^2_{\uparrow}t^2_{\downarrow} \Big(\frac{1}{4} - S^{z}_{\ell}S^{z}_{\ell+1}\Big)+
    \sum_{m} \big(t^2_{\uparrow}-t^2_{\downarrow}\big)^2 \Big(\frac{1}{4} - S^{z}_{2m-1}S^{z}_{2m+1}\Big)+
    \sum_{\ell} \frac{t^4_{\uparrow}-t^4_{\downarrow}}{2}(-1)^\ell S^{z}_{\ell} + \nonumber\\
&& +\sum_{m} 2\big(t^4_{\uparrow}-t^4_{\downarrow}\big)
    \Big(\frac{1}{4}-S^{z}_{2m-1}S^{z}_{2m}\Big)S^{z}_{2m+1}-
    \sum_{\ell} t_{\uparrow}t_{\downarrow}\big(t^2_{\uparrow}+t^2_{\downarrow}\big)
    \big(S^{x}_{\ell}S^{x}_{\ell+1} + S^{y}_{\ell}S^{y}_{\ell+1}\big)  - \nonumber\\
&& -\sum_{m} 2t_{\uparrow}t_{\downarrow}\big(t^2_{\uparrow}-
   t^2_{\downarrow}\big) \big[ \big(S^{x}_{2m-1}S^{x}_{2m} + S^{y}_{2m-1}S^{y}_{2m}\big)S^{z}_{2m+1}+
   S^{z}_{2m-1}\big(S^{x}_{2m}S^{x}_{2m+1} + S^{y}_{2m}S^{y}_{2m+1}\big) \big] \, .
%%%%%%%%%%%%%%%%%%%%%%%%%%%%
\end{eqnarray}
%%%%%%%%%%%%%%%%%%%%%%%%%%%%%%%%%%%%%%%%%%

In a similar manner one can derive the expressions of the operators $T^{e}_{-1}T^{pe}_{0}T^{po}_{0}T^{e}_{1}$, $T^{o}_{-1}T^{do}_{0}T^{de}_{0}T^{o}_{1}$ and $T^{e}_{-1}T^{de}_{0}T^{do}_{0}T^{e}_{1}$, belonging to the same group as the operator considered above. However, it should be noted that one can obtain all of these expressions directly from (\ref{4T-A1-2}) by switching the roles of the odd and the even sites and/or interchanging the hopping amplitudes $t_{\uparrow} \leftrightarrow t_{\downarrow}$\,:
%%%%%%%%%%%%%%%%%%%%%%%%%%%%%%%%%%%%%%%%%%%
\begin{eqnarray}
T^{e}_{-1}T^{pe}_{0}T^{po}_{0}T^{e}_{1} &=&
 T^{o}_{-1}T^{po}_{0}T^{pe}_{0}T^{o}_{1}\,\big(\,2k+1,\, 2m,\, 2n+1 \rightarrow 2k,\, 2m+1,\, 2n\, \big) = \nonumber\\
&=& \sum_{\ell}  2 t^2_{\uparrow}t^2_{\downarrow} \Big(\frac{1}{4} - S^{z}_{\ell}S^{z}_{\ell+1}\Big)+
    \sum_{m} \big(t^2_{\uparrow}-t^2_{\downarrow}\big)^2 \Big(\frac{1}{4} - S^{z}_{2m}S^{z}_{2m+2}\Big)-
    \sum_{\ell} \frac{t^4_{\uparrow}-t^4_{\downarrow}}{2}(-1)^\ell S^{z}_{\ell} + \nonumber\\
&& +\sum_{m} 2\big(t^4_{\uparrow}-t^4_{\downarrow}\big)
    \Big(\frac{1}{4}-S^{z}_{2m}S^{z}_{2m+1}\Big)S^{z}_{2m+2}-
    \sum_{\ell} t_{\uparrow}t_{\downarrow}\big(t^2_{\uparrow}+t^2_{\downarrow}\big)
    \big(S^{x}_{\ell}S^{x}_{\ell+1} + S^{y}_{\ell}S^{y}_{\ell+1}\big) -  \nonumber\\
&& -\sum_{m} 2t_{\uparrow}t_{\downarrow}\big(t^2_{\uparrow}-
   t^2_{\downarrow}\big) \big[ \big(S^{x}_{2m}S^{x}_{2m+1} + S^{y}_{2m}S^{y}_{2m+1}\big)S^{z}_{2m+2}+
   S^{z}_{2m}\big(S^{x}_{2m+1}S^{x}_{2m+2} + S^{y}_{2m+1}S^{y}_{2m+2}\big) \big], \hspace{5mm} \\
T^{o}_{-1}T^{do}_{0}T^{de}_{0}T^{o}_{1}&=&
 T^{o}_{-1}T^{po}_{0}T^{pe}_{0}T^{o}_{1}\,\big(\,2k+1,\, 2m,\, 2n+1 \rightarrow 2k,\, 2m+1,\, 2n \hspace{4mm} \rm{and} \hspace*{4mm} t_{\uparrow} \leftrightarrow t_{\downarrow}\,\big) = \nonumber\\
&=& \sum_{\ell}  2 t^2_{\uparrow}t^2_{\downarrow} \Big(\frac{1}{4} - S^{z}_{\ell}S^{z}_{\ell+1}\Big)+
    \sum_{m} \big(t^2_{\uparrow}-t^2_{\downarrow}\big)^2 \Big(\frac{1}{4} - S^{z}_{2m}S^{z}_{2m+2}\Big)+
    \sum_{\ell} \frac{t^4_{\uparrow}-t^4_{\downarrow}}{2}(-1)^\ell S^{z}_{\ell} - \nonumber\\
&& -\sum_{m} 2\big(t^4_{\uparrow}-t^4_{\downarrow}\big)
    \Big(\frac{1}{4}-S^{z}_{2m}S^{z}_{2m+1}\Big)S^{z}_{2m+2}-
    \sum_{\ell} t_{\uparrow}t_{\downarrow}\big(t^2_{\uparrow}+t^2_{\downarrow}\big)
    \big(S^{x}_{\ell}S^{x}_{\ell+1} + S^{y}_{\ell}S^{y}_{\ell+1}\big) +   \nonumber\\
&& +\sum_{m} 2t_{\uparrow}t_{\downarrow}\big(t^2_{\uparrow}-
   t^2_{\downarrow}\big) \big[ \big(S^{x}_{2m}S^{x}_{2m+1} + S^{y}_{2m}S^{y}_{2m+1}\big)S^{z}_{2m+2}+
   S^{z}_{2m}\big(S^{x}_{2m+1}S^{x}_{2m+2} + S^{y}_{2m+1}S^{y}_{2m+2}\big) \big], \hspace{5mm} \\
T^{e}_{-1}T^{de}_{0}T^{do}_{0}T^{e}_{1} &=& T^{o}_{-1}T^{po}_{0}T^{pe}_{0}T^{o}_{1}\,\big(\,t_{\uparrow} \leftrightarrow t_{\downarrow}\,\big) = \nonumber\\
&=& \sum_{\ell}  2 t^2_{\uparrow}t^2_{\downarrow} \Big(\frac{1}{4} - S^{z}_{\ell}S^{z}_{\ell+1}\Big)+
    \sum_{m} \big(t^2_{\uparrow}-t^2_{\downarrow}\big)^2 \Big(\frac{1}{4} - S^{z}_{2m-1}S^{z}_{2m+1}\Big)-
    \sum_{\ell} \frac{t^4_{\uparrow}-t^4_{\downarrow}}{2}(-1)^\ell S^{z}_{\ell} - \nonumber\\
&& -\sum_{m} 2\big(t^4_{\uparrow}-t^4_{\downarrow}\big)
    \Big(\frac{1}{4}-S^{z}_{2m-1}S^{z}_{2m}\Big)S^{z}_{2m+1}-
    \sum_{\ell} t_{\uparrow}t_{\downarrow}\big(t^2_{\uparrow}+t^2_{\downarrow}\big)
    \big(S^{x}_{\ell}S^{x}_{\ell+1} + S^{y}_{\ell}S^{y}_{\ell+1}\big) +   \nonumber\\
&& +\sum_{m} 2t_{\uparrow}t_{\downarrow}\big(t^2_{\uparrow}-
   t^2_{\downarrow}\big) \big[ \big(S^{x}_{2m-1}S^{x}_{2m} + S^{y}_{2m-1}S^{y}_{2m}\big)S^{z}_{2m+1}+
   S^{z}_{2m-1}\big(S^{x}_{2m}S^{x}_{2m+1} + S^{y}_{2m}S^{y}_{2m+1}\big) \big]. \hspace{5mm}
\end{eqnarray}
%%%%%%%%%%%%%%%%%%%%%%%%%%%%%%%%%%%%%%%%%%

The expressions for the remaining four operators, corresponding to the processes where the pair is created on one site and annihilated on a neighboring one, can be established in an analogous way (utilizing where necessary the freedom of renaming the lattice indices $k \leftrightarrow n$ to facilitate the calculation):
%%%%%%%%%%%%%%%%%%%%%%%%%%%%%%%%%%%%%%%%%%
\begin{eqnarray}
&& \hspace{-12mm} T^{e}_{-1}T^{de}_{0}T^{pe}_{0}T^{o}_{1} =  \nonumber \\
&=&\!\!\sum_{[n,m,k]}\sum_{\sigma}\eta(\sigma)\,t_{\sigma}\,
X_{2n+1}^{\sigma 0}X_{2m}^{\sigmabar 2} \,\cdot
\sum_{\beta}t_{\betabar}\,t^{\phantom{0}}_{\beta}\,X_{2m}^{2\beta}X_{2k+1}^{\beta2}
X_{2m}^{\beta 0}X_{2n+1}^{0\beta}\,\cdot
\sum_{\alpha}\eta(\alpha)\, t_{\alpha}\, X_{2k+1}^{2 \alphabar}X_{2m}^{0 \alpha} \nonumber \\
&=&-\!\!\sum_{[n,m,k]}\,\sum_{\alpha,\beta,\sigma}
\eta(\alpha)\,\eta(\sigma)\,t_{\alpha}\,t_{\sigma}\,t_{\betabar}\,t^{\phantom{0}}_{\beta}\,
X_{2k+1}^{\beta \alphabar}X_{2m}^{\sigmabar \alpha}X_{2n+1}^{\sigma \beta}  \nonumber\\
&=&\!\!\sum_{[n,m,k]} \Big[
   -t^3_{\uparrow}t_{\downarrow}\Big(X_{2k+1}^{\uparrow \downarrow}X_{2m}^{\downarrow \uparrow}X_{2n+1}^{\uparrow \uparrow}+X_{2k+1}^{\downarrow \downarrow}X_{2m}^{\downarrow \uparrow}X_{2n+1}^{\uparrow \downarrow}\Big)
-t_{\uparrow}t^3_{\downarrow}\,\Big(X_{2k+1}^{\uparrow \uparrow}X_{2m}^{\uparrow \downarrow}X_{2n+1}^{\downarrow \uparrow}+X_{2k+1}^{\downarrow
\uparrow}X_{2m}^{\uparrow \downarrow}X_{2n+1}^{\downarrow \downarrow}\Big) + \nonumber\\
&& \hspace{13mm} + \, t^2_{\uparrow}t^2_{\downarrow}\,\Big(X_{2k+1}^{\downarrow \uparrow}X_{2m}^{\downarrow \downarrow}X_{2n+1}^{\uparrow \downarrow} + X_{2k+1}^{\uparrow \downarrow}X_{2m}^{\uparrow \uparrow}X_{2n+1}^{\downarrow \uparrow} + X_{2k+1}^{\downarrow \downarrow}X_{2m}^{\uparrow
\uparrow}X_{2n+1}^{\downarrow \downarrow}+ X_{2k+1}^{\uparrow \uparrow}X_{2m}^{\downarrow \downarrow}X_{2n+1}^{\uparrow \uparrow}\Big)  \Big] \nonumber\\
&=&\!\!\sum_{[n,m,k]} \Big[
   -t^3_{\uparrow}t_{\downarrow}\,X_{2k+1}^{\uparrow \downarrow}X_{2m}^{\downarrow \uparrow}
   -t_{\uparrow}t^3_{\downarrow}\,X_{2k+1}^{\downarrow \uparrow}X_{2m}^{\uparrow \downarrow}
   + t^2_{\uparrow}t^2_{\downarrow}\,\Big(X_{2k+1}^{\uparrow \downarrow}
   X_{2n+1}^{\downarrow \uparrow} +  X_{2k+1}^{\downarrow \downarrow}X_{2m}^{\uparrow
\uparrow}X_{2n+1}^{\downarrow \downarrow}+ X_{2k+1}^{\uparrow \uparrow}X_{2m}^{\downarrow \downarrow}X_{2n+1}^{\uparrow \uparrow} \Big) \Big]  \nonumber\\
&=&\!\!\sum_{[n,m,k]} \bigg[
   -t^3_{\uparrow}t_{\downarrow}\,S_{2k+1}^{+}S_{2m}^{-}
   -t_{\uparrow}t^3_{\downarrow}\,S_{2k+1}^{-}S_{2m}^{+}
   + t^2_{\uparrow}t^2_{\downarrow}\,\Big[S_{2k+1}^{+}S_{2n+1}^{-} + \nonumber \\
&& \hspace{13mm} + \, \Big(\frac{1}{2} - S^{z}_{2k+1}\Big)\Big(\frac{1}{2} + S^{z}_{2m}\Big)
\Big(\frac{1}{2} - S^{z}_{2n+1}\Big) + \Big(\frac{1}{2} + S^{z}_{2k+1}\Big)
\Big(\frac{1}{2} - S^{z}_{2m}\Big) \Big(\frac{1}{2} + S^{z}_{2n+1}\Big) \Big]  \bigg]  \nonumber\\
&=&\!\!\sum_{[n,m,k]} \bigg[
   -t_{\uparrow}t_{\downarrow}\big(t^2_{\uparrow}+t^2_{\downarrow}\big)\big(S_{2k+1}^{x}S_{2m}^{x}+ S_{2k+1}^{y}S_{2m}^{y}\big) + i\,t_{\uparrow}t_{\downarrow}\big(t^2_{\uparrow}-t^2_{\downarrow}\big)\big(S_{2k+1}^{x}S_{2m}^{y} - S_{2k+1}^{y}S_{2m}^{x}\big) +  \nonumber\\
&& \hspace{13mm} + \,  t^2_{\uparrow}t^2_{\downarrow}\,\Big[S_{2k+1}^{+}S_{2n+1}^{-} +
\Big(\frac{1}{4} - S^{z}_{2k+1}S^{z}_{2m} \Big) + \Big(\frac{1}{4} - S^{z}_{2m}S^{z}_{2n+1}\Big) +
\Big( S^{z}_{2k+1}S^{z}_{2n+1} - \frac{1}{4}\Big) \, \Big] \, \bigg] \nonumber\\
&=& -\sum_{\ell}t_{\uparrow}t_{\downarrow}\big(t^2_{\uparrow}+t^2_{\downarrow}\big)
\big(S_{\ell}^{x}S_{\ell+1}^{x}+ S_{\ell}^{y}S_{\ell+1}^{y}\big) + \nonumber \\
&& + \sum_{m} i\, t_{\uparrow}t_{\downarrow}\big(t^2_{\uparrow}-t^2_{\downarrow}\big)
\big[S_{2m-1}^{x}S_{2m}^{y} - S_{2m-1}^{y}S_{2m}^{x}-\big(S_{2m}^{x}S_{2m+1}^{y} - S_{2m}^{y}S_{2m+1}^{x}\big)\big] +  \nonumber\\
&& + \sum_{m} 2 t^2_{\uparrow}t^2_{\downarrow}
\big(S^{x}_{2m-1}S^{x}_{2m+1} + S^{y}_{2m-1}S^{y}_{2m+1}\big) +
\sum_{\ell} 2 t^2_{\uparrow}t^2_{\downarrow}\Big(\frac{1}{4} - S^{z}_{\ell}S^{z}_{\ell+1}\Big)
+ \sum_{m} 2 t^2_{\uparrow}t^2_{\downarrow}\Big(S^{z}_{2m-1}S^{z}_{2m+1}-\frac{1}{4}\Big) \, ,
\end{eqnarray}
\begin{eqnarray}
T^{o}_{-1}T^{do}_{0}T^{po}_{0}T^{e}_{1} &=& T^{e}_{-1}T^{de}_{0}T^{pe}_{0}T^{o}_{1}\,\big(\,2k+1,\, 2m,\, 2m+1 \rightarrow 2k,\, 2m+1,\, 2n \, \big) = \hspace{6cm} \nonumber\\
&=& -\sum_{\ell} t_{\uparrow}t_{\downarrow}(t^2_{\uparrow}+t^2_{\downarrow})
\big(S_{\ell}^{x}S_{\ell+1}^{x}+ S_{\ell}^{y}S_{\ell+1}^{y}\big) +  \nonumber\\
&& + \sum_{m} i\, t_{\uparrow}t_{\downarrow}\big(t^2_{\uparrow}-t^2_{\downarrow}\big)
\big[S_{2m}^{x}S_{2m+1}^{y} - S_{2m}^{y}S_{2m+1}^{x}-\big(S_{2m+1}^{x}S_{2m+2}^{y} - S_{2m+1}^{y}S_{2m+2}^{x}\big)\big] +  \nonumber\\
&& + \sum_{\ell} 2 t^2_{\uparrow}t^2_{\downarrow}\Big(\frac{1}{4} - S^{z}_{\ell}S^{z}_{\ell+1}\Big)
 + \sum_{m} 2t^2_{\uparrow}t^2_{\downarrow}\Big({\bf S}_{2m} \cdot {\bf S}_{2m+2}-\frac{1}{4}\Big) \, , \\
\nonumber \\
%%%%%%%%%%%%%%%%%%%%%%%%%%%%%%%%%%%%%%%%%%%%
%%%%%%%%%%%%%%%%%%%%%%%%%%%%%%%%%%%%%%%%%%%%
T^{o}_{-1}T^{po}_{0}T^{do}_{0}T^{e}_{1} &=& T^{e}_{-1}T^{de}_{0}T^{pe}_{0}T^{o}_{1}\,\big(\,t_{\uparrow} \leftrightarrow t_{\downarrow}\,\big) = \nonumber\\
&=&  -\sum_{\ell}t_{\uparrow}t_{\downarrow}\big(t^2_{\uparrow}+t^2_{\downarrow}\big)
\big(S_{\ell}^{x}S_{\ell+1}^{x}+ S_{\ell}^{y}S_{\ell+1}^{y}\big) -  \nonumber\\
&& -\sum_{m} i\, t_{\uparrow}t_{\downarrow}\big(t^2_{\uparrow}-t^2_{\downarrow}\big)
\big[S_{2m-1}^{x}S_{2m}^{y} - S_{2m-1}^{y}S_{2m}^{x}-\big(S_{2m}^{x}S_{2m+1}^{y} - S_{2m}^{y}S_{2m+1}^{x}\big)\big] +  \nonumber\\
&& +\sum_{\ell} 2 t^2_{\uparrow}t^2_{\downarrow}\Big(\frac{1}{4} - S^{z}_{\ell}S^{z}_{\ell+1}\Big)
+\sum_{m} 2t^2_{\uparrow}t^2_{\downarrow}\Big({\bf S}_{2m-1} \cdot {\bf S}_{2m+1}-\frac{1}{4}\Big) \, , \\
\nonumber \\
%%%%%%%%%%%%%%%%%%%%%%%%%%%%%%%%%%%%%%%%%%%%%
%%%%%%%%%%%%%%%%%%%%%%%%%%%%%%%%%%%%%%%%%%%%%
T^{e}_{-1}T^{pe}_{0}T^{de}_{0}T^{o}_{1} &=&  T^{e}_{-1}T^{de}_{0}T^{pe}_{0}T^{o}_{1}\,\big(\,2k+1,\, 2m,\, 2n+1 \rightarrow 2k,\, 2m+1,\, 2n \hspace{4mm} \rm{and} \hspace*{4mm} t_{\uparrow} \leftrightarrow t_{\downarrow}\,\big) = \nonumber\\
&=& -\sum_{\ell} t_{\uparrow}t_{\downarrow}(t^2_{\uparrow}+t^2_{\downarrow})
\big(S_{\ell}^{x}S_{\ell+1}^{x}+ S_{\ell}^{y}S_{\ell+1}^{y}\big) -  \nonumber\\
&& - \sum_{m} i\, t_{\uparrow}t_{\downarrow}\big(t^2_{\uparrow}-t^2_{\downarrow}\big)
\big[S_{2m}^{x}S_{2m+1}^{y} - S_{2m}^{y}S_{2m+1}^{x}-\big(S_{2m+1}^{x}S_{2m+2}^{y} - S_{2m+1}^{y}S_{2m+2}^{x}\big)\big] +  \nonumber\\
&& + \sum_{\ell} 2 t^2_{\uparrow}t^2_{\downarrow}\Big(\frac{1}{4} - S^{z}_{\ell}S^{z}_{\ell+1}\Big)
 + \sum_{m} 2t^2_{\uparrow}t^2_{\downarrow}\Big({\bf S}_{2m} \cdot {\bf S}_{2m+2}-\frac{1}{4}\Big) \, .
\end{eqnarray}

\end{document}